\documentclass[11pt,twocolumn]{aastex631}
\usepackage{multirow}
\usepackage{makecell}

\newcommand\region{{M17\,SWex}}%
\newcommand\schii{{\sc H\,ii}}%
\newcommand\Msol{{\rm M_{\odot}}}%
\newcommand\um{{$\mu \mathrm{m}$}}%
\newcommand\mpflux{{\mathrm{f_m}}}%
\newcommand\mjybeam{{\rm mJy\,beam^{-1}}}%
\graphicspath{{./}{figures/}}

\begin{document}

\title{Submillimeter and Mid-Infrared Variability of Young Stellar Objects in the \region\ Intermediate-Mass Star-Forming Region}

\author[0000-0001-8467-3736]{Geumsook Park}
\affiliation{Telepix Co., Ltd., 17, Techno 4-ro, Yuseong-gu, Daejeon 34013, Republic of Korea}
\affiliation{Research Institute of Natural Sciences, Chungnam National University, 99 Daehak-ro, Yuseong-gu, Daejeon 34134, Republic of Korea}
\affiliation{Korea Astronomy and Space Science Institute, 776 Daedeokdae-ro, Yuseong-gu, Daejeon 34055, Republic of Korea}

\author[0000-0002-6773-459X]{Doug Johnstone}
\affiliation{NRC Herzberg Astronomy and Astrophysics, 5071 West Saanich Rd, Victoria, BC, V9E 2E7, Canada}
\affiliation{Department of Physics and Astronomy, University of Victoria, Victoria, BC, V8P 5C2, Canada}
 
\author[0000-0003-1894-1880]{Carlos Contreras Pe\~{n}a}
\affiliation{Department of Physics and Astronomy, Seoul National University, 1 Gwanak-ro, Gwanak-gu, Seoul 08826, Korea}
\affiliation{Research Institute of Basic Sciences, Seoul National University, Seoul 08826, Republic of Korea}
 
\author[0000-0003-3119-2087]{Jeong-Eun Lee}
\affiliation{Department of Physics and Astronomy, Seoul National University, 1 Gwanak-ro, Gwanak-gu, Seoul 08826, Korea}
\affiliation{SNU Astronomy Research Center, Seoul National University, 1 Gwanak-ro, Gwanak-gu, Seoul 08826, Korea}

\author{Sheng-Yuan Liu}
\affiliation{Academia Sinica Institute of Astronomy and Astrophysics, 11F of AS/NTU Astronomy-Mathematics Building, No.1, Sec. 4, Roosevelt Rd, Taipei 10617, Taiwan, R.O.C.}

\author[0000-0002-7154-6065]{Gregory Herczeg}
\affiliation{Kavli Institute for Astronomy and Astrophysics, Peking University, Yiheyuan Lu 5, Haidian Qu, 100871 Beijing, Peoples Republic of China}
\affiliation{Department of Astronomy, Peking University, Yiheyuan 5, Haidian Qu, 100871 Beijing, China}

\author[0000-0002-6956-0730]{Steve Mairs}
\affiliation{NRC Herzberg Astronomy and Astrophysics, 5071 West Saanich Rd, Victoria, BC, V9E 2E7, Canada}
\affiliation{East Asian Observatory, 660 N. A`oh\={o}k\={u} Place,
Hilo, Hawai`i, 96720, USA}

\author[0000-0003-0849-0692]{Zhiwei Chen}
\affiliation{Purple Mountain Observatory, Chinese Academy of Sciences, 10 Yuanhua, 210023 Nanjing, China}

\author{Jennifer Hatchell}
\affiliation{Physics and Astronomy, University of Exeter, Stocker Road, Exeter EX4 4QL, UK}
 
\author[0000-0003-2412-7092]{Kee-Tae Kim}
\affiliation{Korea Astronomy and Space Science Institute, 776 Daedeokdae-ro, Yuseong-gu, Daejeon 34055, Republic of Korea}
\affiliation{University of Science and Technology, Korea (UST), 217 Gajeong-ro, Yuseong-gu, Daejeon 34113, Republic of Korea}
 
\author{Mi-Ryang Kim}
\affiliation{Department of Physics and Astronomy, Seoul National University, 1 Gwanak-ro, Gwanak-gu, Seoul 08826, Korea}

\author{Keping Qiu}
\affiliation{School of Astronomy and Space Science, Nanjing University Xianlin Campus, 163 Xianlin Avenue, Qixia District, Nanjing, Jiangsu, China, 210023}
 
\author{Yao-Te Wang}
\affiliation{Academia Sinica Institute of Astronomy and Astrophysics, 11F of AS/NTU Astronomy-Mathematics Building, No.1, Sec. 4, Roosevelt Rd, Taipei 10617, Taiwan, R.O.C.}
\affiliation{Graduate Institute of Astrophysics, National Taiwan University, No. 1, Sec. 4, Roosevelt Rd., Taipei 10617, Taiwan, R.O.C.}
 
\author{Xu Zhang}
\affiliation{School of Astronomy and Space Science, Nanjing University Xianlin Campus, 163 Xianlin Avenue, Qixia District, Nanjing, Jiangsu, China, 210023}

\author{The JCMT Transient Team}

\begin{abstract}

We present a comprehensive analysis of young stellar object (YSO) variability within the M17 Southwest Extension (\region), using 3.5 years of monitoring data from the JCMT Transient Survey at sub-millimeter (sub-mm) and 9 years from the NEOWISE mission at mid-infrared (mid-IR).
Our study encompasses observations of 147 bright sub-mm peaks identified within our deep JCMT co-added map as well as 156 YSOs in NEOWISE W1 and 179 in W2 that were previously identified in Spitzer surveys.
We find three robust sub-mm variables: two are candidate YSOs and one is a likely extragalactic source.
At mid-IR wavelengths, our analysis reveals 
secular and stochastic variability in 47 YSOs, with the highest fraction of secular variability occurring at the earliest evolutionary stage. This is similar to what has previously been observed for low-mass YSO variability within the Gould Belt. However, we observe less overall variability in \region\ at both the sub-mm and mid-IR. We suspect that this lower fraction is due to the greater distance to \region.
Our findings showcase the utility of multi-wavelength observations to better capture the complex variability phenomena inherent to star formation processes and demonstrate the importance of years-long monitoring of a diverse selection of star-forming environments.

\end{abstract}


\section{Introduction} \label{sec:intro}

Variability in young stellar objects (YSOs) provides a powerful tool for understanding the physical processes that occur during star formation and the history of mass assembly onto the forming star
(see review by \citealt{fischer2023} and studies by, e.g., \citealt{carpenter2001, caramazza2007, billot2012, cody2014, guarcello2017}).
While optical and near-infrared (near-IR) surveys  have revealed the complex variability of the less obscured disk-hosting and diskless YSOs, those protostars that are deeply embedded within their nascent envelopes are much more challenging to monitor.
The optical depth and geometric complexities, such as the envelope structures and outflow cavities, significantly impact observed fluxes, including at mid-infrared (mid-IR) wavelengths \citep[e.g.][]{whitney2003}.
These complications have a significant influence on our interpretation of mid-IR outbursts from YSOs, potentially making our understanding of their early evolution stages more uncertain \citep[e.g.,][]{fischer2019, Macfarlane2019a, Macfarlane2019b, baek2020, fischer2024}.
This complexity is significantly reduced by far-infrared (far-IR) through submillimeter (sub-mm) observations.
Because these wavelengths respond to the reprocessing of the protostellar luminosity by the enshrouding envelope, they 
are mostly robust to geometrical uncertainties and help to clarify the underlying nature of the variability and the mechanisms behind it \citep{johnstone2013,fischer2024}.

Variability in the accretion rate of mass onto stars manifests as a change in the protostellar luminosity as released gravitational energy heats the infalling matter. Thus a history of the mass assembly of stars is directly observable through variations in the protostellar brightness.  Accretion variability, especially that associated with intense bursts, may play an important role in the mass assembly of the protostar \citep{scholz2013, fischer2019, w.park2021}, impact the chemical evolution of both the envelope and the disk surrounding the YSO \citep[e.g.][]{lee07,jorgensen15,molyarova18,hsieh19} and potentially affect the contraction of the forming star \citep[e.g.][]{baraffe10,hosokawa11,kunitomo17}.
As mentioned, optically thick YSO envelopes obscure this variability at optical through near-IR wavelengths during the early stages of stellar evolution.
These are, however, the phases where stars acquire most of their mass, emphasizing the importance of quantifying the role of variability during this epoch.

The best wavelength range to monitor accretion variability from deeply embedded protostars is the far-IR. Unfortunately this is not possible from the ground. Sub-mm monitoring of nearby low-mass star-forming regions by the JCMT Transient Survey \citep{herczeg2017} since December 2015 has allowed an exploration of accretion variability properties for a modest ensemble of deeply embedded protostars during their primary growth spurts \citep{y.lee2021,Mairs2024}.  Several sources with intriguing sub-mm light curves have been further analyzed using multi-wavelength data sets and sub-mm interferometry, including EC53 in Serpens Main \citep[also known as V371 Ser,][]{y.lee2020,baek2020, francis22} and HOPS 373 in Orion \citep{Yoon2022, jelee2023, shlee2024}.

Furthermore, despite the known uncertainties in connecting mid-IR variability directly to accretion variations, the JCMT Transient team has found excellent agreement between the long-term sub-mm and mid-IR variability for specific {sub-mm sources \citep{Carlos2020}.  
Statistical analyses of mid-IR variability also indicate similar fractions as in the sub-mm of long-term variability, likely due to accretion 
(\citealt{w.park2021}, see also analysis by \citealt{zakri22}). Thus, while for any individual YSO one must use caution when interpreting the mid-IR lightcurve, for long-term monitoring over years, 
we find that sources showing years-long sub-mm variations also have similar mid-IR variability on similar timescales.}
Still, while there is evidence for a rough calibration between the amplitude of mid-IR variations and the underlying protostar luminosity change \citep[see for example,][]{Carlos2020}, for individual sources this can be strongly affected by the presence of mid-IR emission lines from unrelated shocks, as seen {in W2} for HOPS 373 by \citet{Yoon2022}.
These findings have crucial implications for theories of star formation, particularly regarding the mechanisms through which stars accumulate their mass during the earliest phases of development \citep[see review by][]{fischer2023}.

The question remains, however, whether similar accretion-driven luminosity bursts occur in more active star-forming regions, such as those capable of forming more massive stars, and how these processes might differ in frequency, amplitude, and duration as a function of environment.
In this context, the M17 Southwest Extension (\region) offers a powerful laboratory for studying star formation processes due to its proximity to the vibrant \schii\ region M17, its status as one of the Galaxy's most prominent infrared dark cloud complexes, and its rich population of YSOs.
Situated at a distance of $\sim 1.8$ kpc \citep{wu2014}, \region ~harbors a diverse assembly of star-forming potential, including 94 Class0/I objects, 179 Class II objects, and 48 objects with ambiguous classification based on the Spitzer GLIMPSE and MIPSGAL survey \citep{povich2010}.
This large population of YSOs across a wide-range of evolutionary stages underscores the region's dynamic environment, appropriate for investigating the early stages of stellar development.
Notably, while \region's proto-OB association currently lacks very massive stars ($> 20 \Msol$), it contains massive cores that are likely to form O stars in the future \citep{povich2010, povich2016}, indicating ongoing massive star formation.
\region\ exhibits dense filamentary structures and is affected by well-aligned magnetic fields perpendicular to these filaments \citep{sugitani2019}. Furthermore, recent studies of the molecular gas content confirm the spatial and kinematic physical connection between M17 and \region\ \citep{shimoikura2019, nguyen2020}.

Our multi-epoch sub-mm monitoring is part of the extended JCMT Transient Survey \citep{herczeg2017, Mairs2024}. 
Combined with mid-IR data acquired during the NEOWISE mission, we are able to infer and analyze a broad range of variable YSO candidates {\citep[e.g.][]{w.park2021}.
Here our primary objective is to examine the nature of YSO variability across sub-mm and mid-IR wavelengths in the dynamic setting of the massive star-forming region \region.

This paper comprehensively analyzes YSO variability across \region, focusing on observations made at sub-mm and mid-IR wavelengths.
Section~2 outlines the observational data employed in this study, including the JCMT Transient Survey and the NEOWISE datasets.
In Section~3, we describe the analytical methods used to identify variable sources.
Section~4 discusses our findings on the characteristics of variability observed in sub-mm and mid-IR wavelengths within \region, including a brief comparative analysis of these results against the variability observed in low-mass star-forming regions.
Finally, Section~5 summarizes our main findings and conclusions.

\section{Observational Data} \label{sec:obs}

\subsection{JCMT Transient Survey} \label{sec:obs_submm}

The primary aim of the JCMT Transient Survey is to measure sub-mm variability of protostars  \citep{herczeg2017, Mairs2024} using the Submillimetre Common User Bolometer Array~2 (SCUBA-2) instrument \citep{holland2013} on the James Clerk Maxwell Telescope (JCMT) in Hawaii.  The survey initially monitored sub-mm continuum emission in eight star-forming regions, all located in the Gould Belt and within 500~pc of the Sun.  In 2020, the program expanded to also include monitoring of fields within four massive star-forming regions (DR21, M17, \region, and S255).  These new targets were chosen to include regions with the potential to form a large number of stars, including some very massive stars, while still being relatively nearby, located at distances of $\lesssim$ 2~kpc.




Observations are conducted simultaneously at 450 and 850~\um\ with effective beam sizes of 9\farcs8 and 14\farcs6, respectively \citep{Dempsey2013}. 
For the JCMT Transient Survey, each star-forming region is observed using the PONG1800 mode \citep{kackley2010}.
This mode scans a circular field with a diameter of $ 30$~arcmin, providing uniform background noise across the mapped area. In practice, a somewhat larger $40$~arcmin diameter field has sufficient noise uniformity for our analysis.
The monitoring cadence for each region is monthly, as long as the region is observable by SCUBA2.
The observing time is set by the weather conditions at the telescope in order to maintain a uniform measurement uncertainty $\sim 10\,\mjybeam$ at 850~\um\ across epochs. Thus, the 450~\um\ observations are only occasionally of high enough quality for measurements.

After data acquisition and running of the default JCMT data reduction pipeline, the JCMT Transient Survey utilizes custom image alignment and relative calibration techniques to ensure reliable light curves.
Details of the observing methods and reduction techniques can be found in papers by \citet{mairs2017_843, Mairs2024}, where it is shown that {relative} calibration between epochs at 850~\um\ is now close to 1\% for Gould Belt regions with roughly 50 measured epochs. 

This study focuses on the newly added 850~\um\ monitoring data set of star-forming region \region, calibrated using the updated \citet{Mairs2024} procedures.
The central position of the observed area for \region\ is located at (R.A., Dec.) $=$ ($18^\mathrm{h}18^\mathrm{m}30^\mathrm{s}$, $-16^\mathrm{d}51^\mathrm{m}24^\mathrm{s}$).
This region's individual epoch observation dates, weather conditions,  quality of each map, and the calibration requirement are provided in Table~\ref{tab:obsjcmt}. 
During our analysis, we discovered that one epoch, taken on 2022-06-28 with a relatively high noise level, was impacted by artifacts such as echoes in the image.
The cause of these issues remains uncertain and may have been due to either observational problems or issues at the telescope on that night.
To ensure the validity and accuracy of our results, we exclude this epoch from our analysis.
The final dataset for this paper consists of 23 epochs of SCUBA-2 850~\um\ imaging of \region\ from February 2020 until August 2023.



\begin{deluxetable}{ccccl cc}
\tabletypesize{\scriptsize}
\setlength{\tabcolsep}{1.mm}
\tablecaption{JCMT 850~\um\ Observational Data Summary \label{tab:obsjcmt}}
\tablehead{
\colhead{Epoch} & \colhead{Date} & \colhead{MJD\tablenotemark{\scriptsize a}} & \colhead{Scan} & \colhead{\phn$\tau_{225}$\tablenotemark{\scriptsize b}} & 
\colhead{Noise} & \colhead{\phn FCF\tablenotemark{\scriptsize c}} \\
& \colhead{(yyyy-mm-dd)} & & & & \colhead{($\rm mJy\,beam^{-1}$)}
}
\startdata
 \phn1   & 2020-02-22                                      & 58901.73 & 74 & 0.036 &  8.2     & 1.179\\
 \phn2   & 2020-05-21                                      & 58990.48 & 35 & 0.116 & 11.9\phn & 0.984\\
 \phn3   & 2020-06-23                                      & 59023.46 & 34 & 0.1   & 11.6\phn & 0.939\\
 \phn4   & 2020-07-30                                      & 59060.35 & 29 & 0.055 &  9.4     & 1.015\\
 \phn5   & 2020-09-02                                      & 59094.17 & 16 & 0.048 &  7.6     & 1.0\phn\phn\\
 \phn6   & 2020-10-10                                      & 59132.21 & 15 & 0.099 & 13.9\phn & 0.845\\
 \phn7   & 2021-03-03                                      & 59276.75 & 84 & 0.041 &  9.3     & 1.037\\
 \phn8   & 2021-04-06                                      & 59310.55 & 46 & 0.05  &  8.8     & 1.038\\
 \phn9   & 2021-05-17                                      & 59351.56 & 75 & 0.048 & 12.6\phn & 0.998\\  
     10  & 2021-06-14                                      & 59379.39 & 30 & 0.056 &  7.8     & 1.053\\
     11  & 2021-08-04                                      & 59430.21 & 15 & 0.114 & 16.2\phn & 0.893\\
     12  & 2021-09-08                                      & 59465.28 & 24 & 0.106 & 13.6\phn & 0.963\\
     13  & 2021-10-08                                      & 59495.18 & 21 & 0.062 & 11.9\phn & 0.948\\
     14  & 2022-03-03                                      & 59641.67 & 53 & 0.071 &  9.1     & 1.102\\
     15  & 2022-05-23                                      & 59722.55 & 37 & 0.049 &  7.1     & 1.047\\
     16  &\phn2022-06-28\tablenotemark{\scriptsize d}      & 59758.36 & 48 & 0.055 & 18.5\phn & 0.752\\
     17  & 2022-07-29                                      & 59789.24 & 11 & 0.076 & 10.4\phn & 1.019\\
     18  & 2022-08-27                                      & 59818.27 & 30 & 0.054 &  9.2     & 1.006\\
     19  & 2022-10-01                                      & 59853.23 & 16 & 0.062 & 10.3\phn & 0.964\\
     20  & 2023-03-28                                      & 60031.60 & 15 & 0.099 & 11.0\phn & 1.040\\
     21  & 2023-05-01                                      & 60065.54 & 34 & 0.106 & 10.7\phn & 1.078\\
     22  & 2023-06-06                                      & 60101.54 & 54 & 0.079 & 11.7\phn & 0.964\\
     23  & 2023-07-15                                      & 60140.30 & 21 & 0.058 & 11.6\phn & 1.029\\
     24  & 2023-08-14                                      & 60170.30 & 12 & 0.062 &  8.8     & 0.951\\
\enddata
\tablenotetext{a}{Modified Julian Day (MJD) is converted from Julian Day by subtracting 2,400,000.5 from the JD value.}
\tablenotetext{b}{The average 225~GHz zenith opacity measured during observations.}
\tablenotetext{c}{The Flux Calibration Factor (FCF) represents the factor derived during the relative calibration step and applied to the observatory-defined default calibration factor \citep{Mairs2024}.}
\tablenotetext{d}{This epoch is not used in our analysis due to poor image quality.}
\end{deluxetable}

\subsection{WISE/NEOWISE} \label{sec:obs_IR}

From January through September 2010, the Wide-field Infrared Survey Explorer \citep[WISE,][]{wright2010} surveyed the entire sky using four mid-infrared (mid-IR) bands at 3.4~\um\ (W1), 4.6~\um\ (W2), 12~\um\ (W3), and 22~\um\ (W4) with angular resolutions of 6\farcs1, 6\farcs4, 6\farcs5, and 12\arcsec, respectively.
An additional four months of operation followed, utilizing only the short wavelength bands of W1 and W2 during the post-cryogenic mission of the Near-Earth Object WISE (NEOWISE) program \citep{mainzer2011}.
The WISE spacecraft was placed in hibernation in February 2011 and was restarted for the NEOWISE Reactivation mission in September 2013 to survey the entire sky for near-Earth objects (NEOs), including comets and asteroids, using only the W1 and W2 bands \citep{mainzer2014}. {NEOWISE continues to operate, with twice yearly all-sky observations. The data release used here contains observations through December 2022\footnote{\url{https://wise2.ipac.caltech.edu/docs/release/neowise/neowise_2023_release_intro.html}}.  For each area of the sky, {\it WISE} takes 10-20 exposures across a few days, with epochs separated by half a year.}


\section{Analysis} \label{sec:analy}

\subsection{A sub-mm source catalog}\label{sec:analy_submm}

We employed the FellWalker algorithm \citep{berry2015} to identify localized sub-mm sources in the co-added image at 850~\um, generated from the observations of \region\ during the first $\sim 1.5$ years. Sources are only included in our catalog if they were located within a 20\arcmin\ radius of the map center, thereby excluding the noisy map edges.
The mean peak flux of each source was then derived from the full co-adedd image (Figure~\ref{fig:JCMT_850_coadd}), which combined data from all 23 epochs.

{We identify 663 sub-mm sources that have an average peak brightness greater than 13~$\mjybeam$ (see Appendix Table~\ref{tab:all_submm}).}  Of these sources, 150 have a mean peak flux above 100~$\mjybeam$, which is about ten times the mean epoch root-mean-square (RMS) noise ($\sim 11~\mjybeam$; see Table~\ref{tab:obsjcmt}), sufficient for secular variability analysis.
Three of these bright sources are excluded because they are located near the 20\arcmin\ radius edge and extend beyond it, introducing increased uncertainty in the peak position and brightness.

Our final catalog for in-depth variability analysis consists of 147 targets brighter than 100 ~$\mjybeam$.  
The original list of almost 700 identified sub-mm sources is also investigated through a statistical analysis of the flux variability, as devised by \citet{johnstone2018}.
For these fainter sources, however, only large variable brightness events are robustly observable.

\begin{figure*}
    \centering
    \includegraphics[width=.7\textwidth]{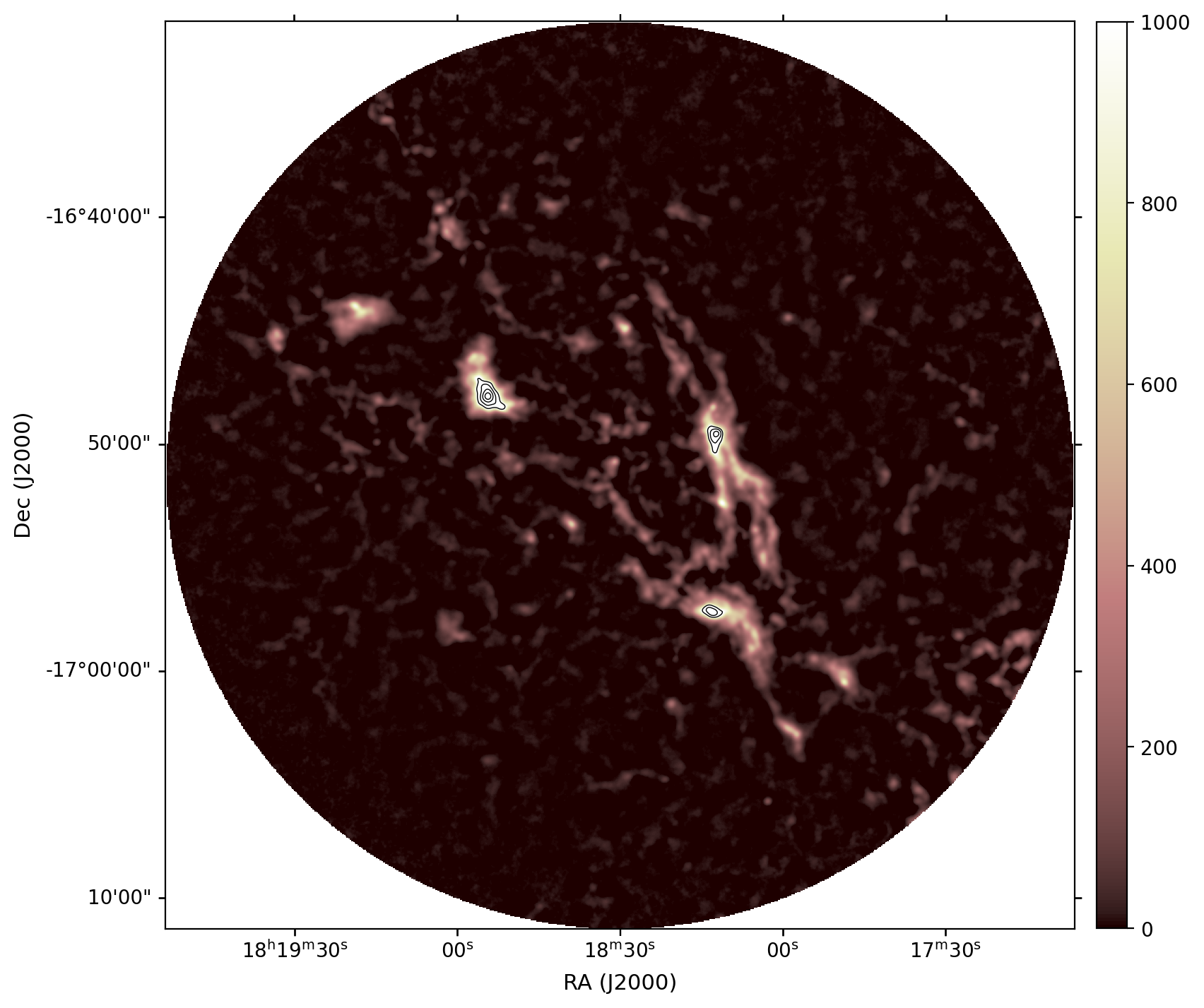}
    \caption{Co-added 850~\um\ image of \region\ using 23 epochs and with contours overlaid at (1100, 2000, 4000, 9000) $\mjybeam$.}
    \label{fig:JCMT_850_coadd}
\end{figure*}

\subsection{Searching for sub-mm Variables}

We evaluate variability of each source by adopting 
 the fiducial standard deviation ($\mathrm{SD_{fid}}$), as implemented by \citet{johnstone2018}.
The $\mathrm{SD_{fid}}$ for a given source $i$, with mean peak flux $\mpflux$, is computed as 
\begin{equation}
    {\mathrm{SD}_\mathrm{fid}}(i) = \sqrt{\left(\mathrm{n}_\mathrm{RMS}\right)^2 + \left({\mathrm{RFCF}_\mathrm{unc}} \times {\mpflux(i)}\right)^2}, 
\end{equation}
where $\mathrm{n_{RMS}}$ represents the typical RMS noise measured across all epochs, primarily affecting faint sources.
We determined $\mathrm{n_{RMS}}$ to be $14~\mjybeam$ for our target region.\footnote{We estimate the typical RMS noise ($\mathrm{n_{RMS}}$) associated with source peak flux measurements by requiring the typical calculated standard deviation for faint sources, those with peak flux  $< 100~\mjybeam$, to be similar to $\mathrm{SD}_\mathrm{fid}$. 
This noise level is somewhat larger than the typical mean RMS map noise found for each epoch because source measurements are affected by additional factors, such as data reduction processing and recovery within local extended emission.
}
The term $\mathrm{RFCF_{unc}}$ denotes the expected relative flux calibration uncertainty between epochs \citep[for more details see][]{johnstone2018, Mairs2024}.
This term predominately impacts bright sources, and we adopt $\mathrm{RFCF_{unc}} = 0.015$, corresponding to a 1.5\% flux calibration uncertainty.
Using the current calibration pipeline, \citet{Mairs2024} find an uncertainty of only 1\% for Gould Belt regions.
Our use of 1.5\% incorporates the additional uncertainty related to the smaller number of observations undertaken to date for \region\ as well as the large scale complexity of the sub-mm emission within the region.
Lastly, $\mpflux$ signifies the mean peak flux of source $i$.

Using the fiducial standard deviation model, we identify outlier sub-mm sources, as shown in Figure~\ref{fig:submm_analysis}a.
This figure plots the measured standard deviation of the source peak flux divided by the fiducial model against the mean peak flux for each sub-mm source.
The upper right quadrant bounded by the two dashed lines of Figure~\ref{fig:submm_analysis}a selects sources that are both bright ($ > 100\,\mjybeam$) and with significant variability ($\mathrm{SD}/\mathrm{SD_{fid}} > 2.5$).
Only one source lies well within the region and is robustly detected as a variable (see Table \ref{tab:variables_submm}). We further note that there are no obvious strong outliers among the many fainter sources (upper left quadrant).

Following the method described by \citet{johnstone2018}, we also searched for sub-mm sources varying in brightness linearly with time, secular variables.
We used the least-squares linear fit method, implemented in the function `scipy.optimize.curve\_fit()', to derive a linear model fit $f_l(i,t)$ for each source $i$ as a function of time $t$, measured from the first epoch $t_0$.
The slope $S(i)$, representing the fractional flux change per year, was obtained by using the following equation:
\begin{equation}
    f_l(i, t) = f_0(i) \left( 1 + S(i) \times (t - t_0)\right),
\end{equation}
where $f_0(i)$ is the initial flux value of the source, and $t_0$ is the time of the first epoch.
We also calculated the uncertainty in the slope $S(i)$ using the `curve\_fit' function, with the expectation that sources with small fractional slope uncertainties indicate robust secular variables.  
In Figure~\ref{fig:submm_analysis}b, we find two bright sources ($ > 100\,\mjybeam$) with $|S/\Delta S| \geq 3$, where $\Delta S$ is the uncertainty in the slope. Neither of these secular sources show significantly enhanced standard deviations (i.e.\ they do not lie in the upper right quadrant in Figure~\ref{fig:submm_analysis}a). Similarly, the variable found to have a significantly enhanced standard deviation in Figure~\ref{fig:submm_analysis}a does not have a clear secular variation suggesting that its a more stochastic sub-mm source.

\begin{figure}
    \centering
    \includegraphics[width=0.5\textwidth]{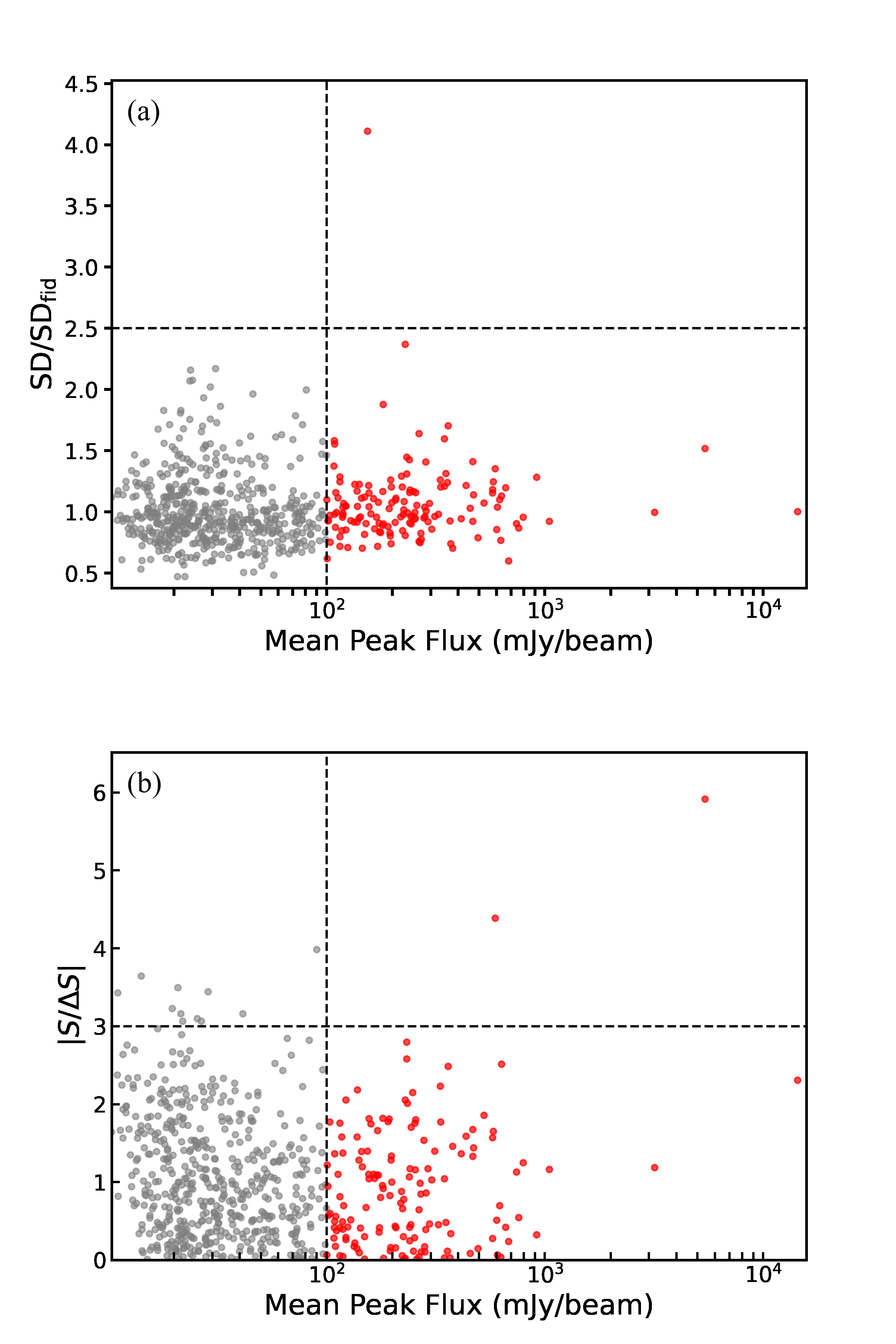}
    \caption{
    Scatter plots of mean peak flux at 850~$\mu$m\ against (a) the standard deviation of the peak flux normalized by the fiducial model and (b) the absolute value of slope $S$ ratioed to its uncertainty $\Delta S$ for sub-mm sources identified in \region.
    Sources with a mean peak flux greater than $100~\mjybeam$ are shown in red, while all others are in grey.
    The dashed lines in each panel indicate thresholds used to identify variable candidates. 
    The dashed vertical line indicates a mean peak flux of 100 $\mjybeam$, the threshold for sources bright enough for statistical significance.
    The dashed horizontal line in Panel (a) indicates 2.5 times the fiducial expectation, a threshold for reliability for stochastic variability. 
    The dashed horizontal line in Panel (b) indicates where $|S/\Delta S| = 3$, which is the threshold for reliability for linear variability as defined by \citet{johnstone2018}. 
    The sources that are marked in red in the upper right corners are considered to be robust variables. 
    There is one robust stochastic variable in (a) and there are two robust linear secular variables in (b).}
    \label{fig:submm_analysis}
\end{figure}

{Figure~\ref{fig:lc_submm} shows the light curves of the three identified robust variables.} We present the sky locations, potential known matches, and the statistical analysis results for the three sub-mm outliers in Table~\ref{tab:variables_submm} and further examine them in Section~\ref{sec:disc_submm}.

\begin{deluxetable*}{ccccc lll}
\tabletypesize{\scriptsize}
\setlength{\tabcolsep}{1pt}
\tablecaption{Statistics of Potential Variable Submillimeter Sources \label{tab:variables_submm}}
\tablehead{
\colhead{} & \colhead{} & \colhead{} & {} & {} & 
\multicolumn{2}{c}{Type} & {} \\
\cline{5-7}
\colhead{ID} & \colhead{$\mpflux$} & \colhead{SD} & \colhead{SD/SD$_{\rm fid}$} & \colhead{$|S/\Delta S|$} & 
\colhead{Source} & \colhead{Variable} & \colhead{Nearby Known Source\tablenotemark{\scriptsize a}} \\
\colhead{(1)} & \colhead{(2)} & \colhead{(3)} & \colhead{(4)} & \colhead{(5)} & 
\colhead{(6)} & \colhead{(7)} & \colhead{(8)} 
}
\startdata
JCMTPP\_J181812.2$-$164933 & 5409.12\phn & 125.06\phn & 1.52 & 5.92 & Star-forming  & Secular    & SPICY~80147, PW2010\_293 (4.9\arcsec)\tablenotemark{\scriptsize b} \\
JCMTPP\_J181803.7$-$165500 &  590.95     &  22.43     & 1.35 & 4.39 & Star-forming  & Secular    & AGAL014.131-00.522\_S (2.0\arcsec)\tablenotemark{\scriptsize c}        \\
JCMTPP\_J181802.8$-$170545 &  153.59     &  58.34     & 4.11 & 1.40 & Quasar        & Stochastic & NVSS~J181802-170542 (3.4\arcsec), IVS~B1851-171, LQAC\_274-017\_001 (4.2\arcsec)\tablenotemark{\scriptsize d}          \\  
\enddata
\tablecomments{ 
(1) Source ID. JCMTPP stands for JCMT Peak Position, followed by Right Ascension and Declination in h:m:s and d:m:s, respectively.;  
(2) Mean peak flux in $\mjybeam$; 
(3) Standard deviation in $\mjybeam$; 
(4) The ratio of the standard deviation to the fiducial standard deviation for each source; 
(5) The absolute value of the slope to its uncertainty ratio; 
(6) Type of source from literature; 
(7) Type of variable identified by this work; 
(8) Known sources that are likely to be associated with the given submm source.\\
}
\tablenotetext{a}{Known sources located within a distance of 6\arcsec. 
The values in parentheses are the distance between the submm peak position and each source.}
\tablenotetext{b}{YSO cataloged by \citet[][PW2010]{povich2010} and \citet[][SPICY]{kuhn2021}.}
\tablenotetext{c}{{Protostellar clump cataloged by ATLASGAL
\citep{Schuller+2009,Contreras+2013,Urquhart+2014}.}}
\tablenotetext{d}{Extragalactic source.
Nearby known sources are listed in order as they appear in the text: \citet{condon1998}, \citet{petrov2011}, and \citet{petrov2011}.}
\end{deluxetable*}

\subsection{Searching for Mid-IR Variables} \label{sec:analy_IR}

Our investigation into mid-IR flux variability of known YSOs in \region\ draws upon the comprehensive catalogs by \citet{povich2010} and \citet{kuhn2021}.
\citet{povich2010} conducted a thorough analysis of archival Spitzer GLIMPSE and MIPSGAL surveys, combined with Two Micron All Sky Survey (2MASS) and Midcourse Space Experiment (MSX) data, to identify 488 candidate YSOs within \region.
This catalog provides a detailed baseline for understanding the initial mass function and the evolutionary stages of YSOs within the cloud complex.
\citet{kuhn2021} expanded on this work by presenting a list of $\sim 120,000$ Spitzer/IRAC candidate YSOs across the Galactic midplane, which included the \region\ region.
Their classification scheme benefited from statistical learning methods tailored to Spitzer's mid-IR capabilities, offering a refined analysis of each objects' nature and the ensemble spatial distribution.
Both catalogs employ a classification scheme that identifies YSOs based on their infrared excess, indicative of circumstellar disks or envelopes, which are key markers of early stellar evolution.
These works provide a critical framework for our study, offering a catalog of YSOs whose variability we aim to explore in the mid-IR.
By using these detailed classifications, we can better understand the mid-IR flux variability in relation to the evolutionary stages of YSOs, from deeply embedded Class 0/I objects to more evolved Class II disks and then {Class III sources}, within the dynamic environment of \region.
Figure~\ref{fig:IR_flowchart} presents a summary flow chart illustrating the mid-IR sample selection process and variability type classification described below.

To allow for direct comparison with the sub-mm observations, we limited the sample of YSOs to those within the boundary of the $40$\arcmin\ diameter JCMT field as described in Section~\ref{sec:analy_submm}. 
Within this area, we identify 583 YSOs from the catalogues.
These YSOs are predominantly at an early evolutionary stage, with  34.1\% {Class 0/I} and 50.5\% {Class II}, suggesting that the observed \region\ is still at an early stage of star formation.  The sample contains only a minor proportion of {Class III sources} (2.7\%), though this is likely due to the complexity of identifying them in distant, dust-obscured, regions. For completeness, ambiguous sources, which cannot be assigned robustly to individual evolutionary classes, make up 12.7\% of the 583 YSOs.

To perform our search for mid-IR counterparts, for each source, we first queried the WISE and NEOWISE single exposure catalogues from the NASA/IPAC Infrared Science Archive (IRSA), using a 3\arcsec\ radius. Then, from all the detections, we determined the average values for the right ascension and declination, and selected single exposures that were located within 2$*sd_{d}$ from the mean location (where $sd_{d}$ is the standard deviation of the distance from the mean location). The next step was to group all of the exposures that were carried out within a few days of each other, as these represent a specific epoch. From these groups, we discarded the brightest and faintest 15\% of the data.  Using the remaining 70\% of the data we estimate the mean MJD, mean magnitude, mean error, and the standard deviation (in magnitudes) of the exposures for the YSO in the epoch. The measurement error was  then calculated by adding, in quadrature, the mean error and the standard deviation within each epoch. This method then provides one epoch of observations every $\sim$ 6 months.

\begin{figure*}
    \centering
    \includegraphics[width=\textwidth]{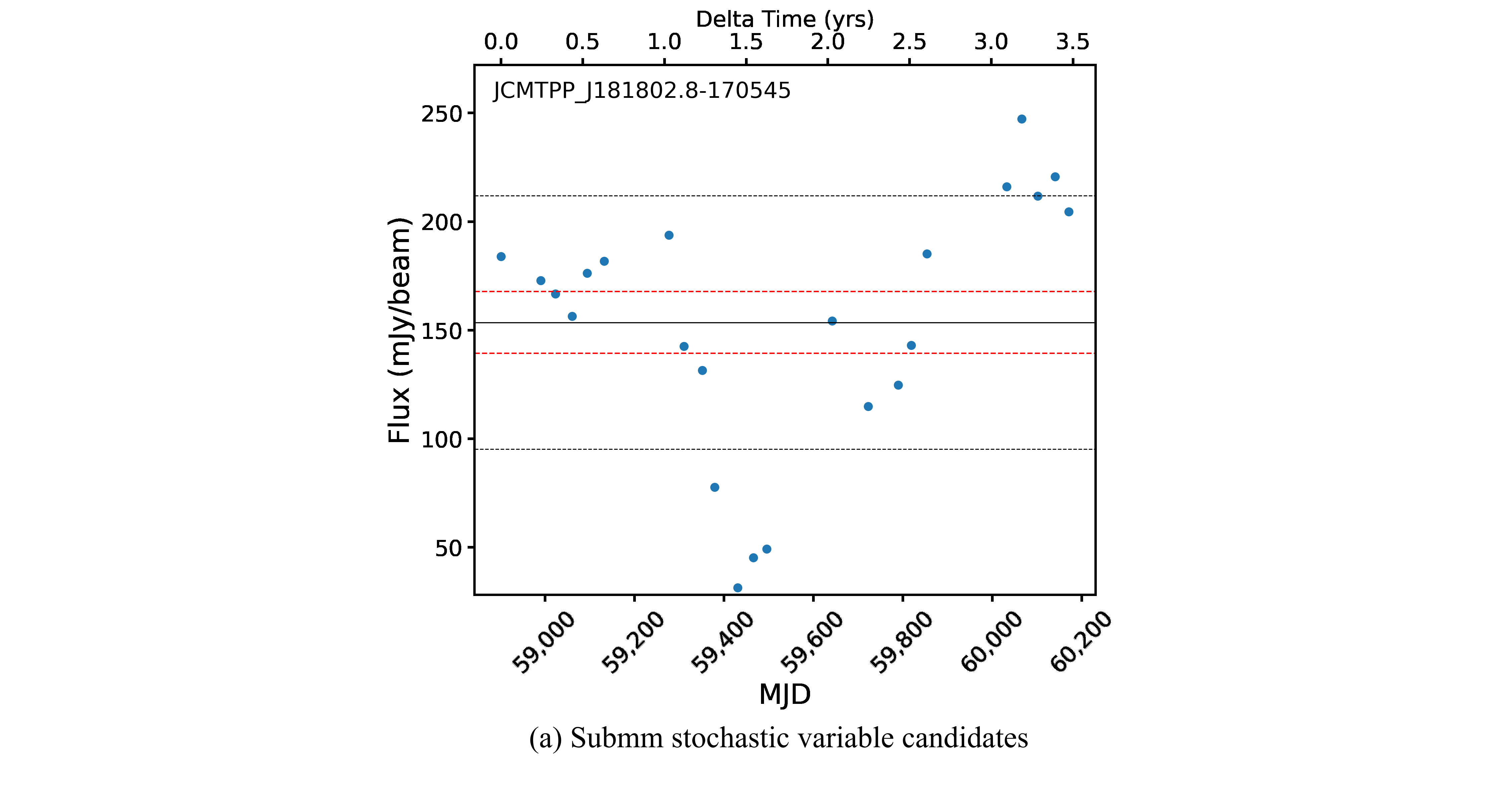}
    \includegraphics[width=\textwidth]{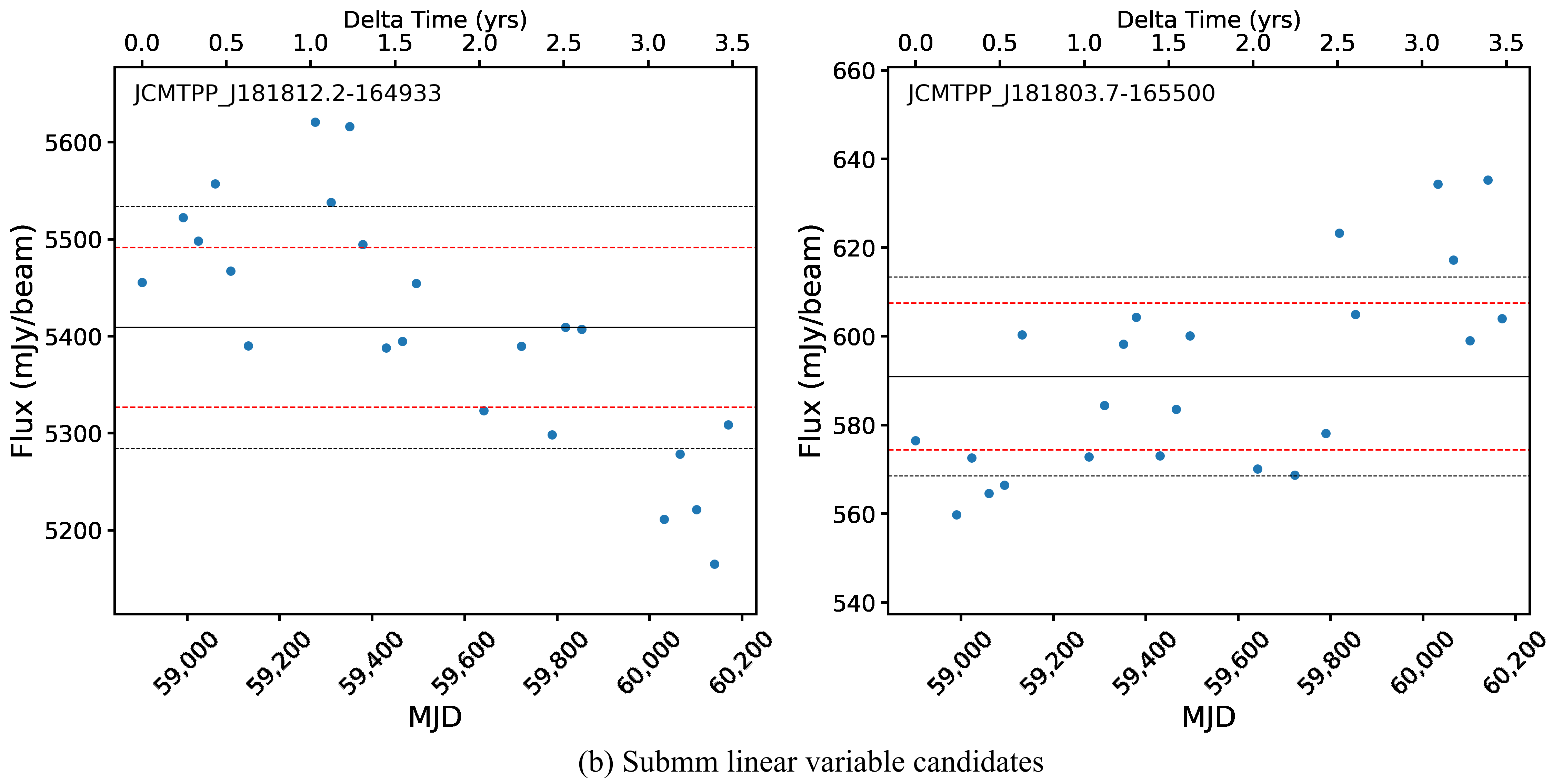}
    \caption{The light curve of potential stochastic (a) or linear (b) variable sub-mm candidates based on the 23 epochs of JCMT observations.
    The black solid and dotted lines represent the mean and $\pm 1\sigma$ standard deviations, while the red dotted lines show the expected fiducial standard deviations.
    The upper x-axis indicates the time relative to the first epoch.
    }
    \label{fig:lc_submm}
\end{figure*}

To analyse variability in the sample and ensure high-quality data, we selected sources with a minimum of 12 epochs for which there are mid-IR measurements at both W1 and W2, and with a mean uncertainty $\mathrm{\sigma(mag)}<0.2$~mag. Among the 583 known YSOs in \region, 185 met the observation threshold with at least 12 epochs. 
Applying as well the measurement uncertainty requirements, we identified 156  YSOs at W1 and 179 at W2. Figure~\ref{fig:IR_flowchart} records the numbers of sources recovered throughout the sample selection and variability classification process. 
This mid-IR-detected sample is dominated by {Class 0/I and Class II}sources, which together make up 95\% of the ensemble, a slightly larger fraction of early stage sources than in the original sample of 583.
Figure~\ref{fig:IR_YSO_sample}a displays the distribution of the known YSO classes and the mean W1 and W2 magnitudes of the final sample objects selected for light curve analysis.


\begin{figure*}
    \centering
    \includegraphics[scale=.5]{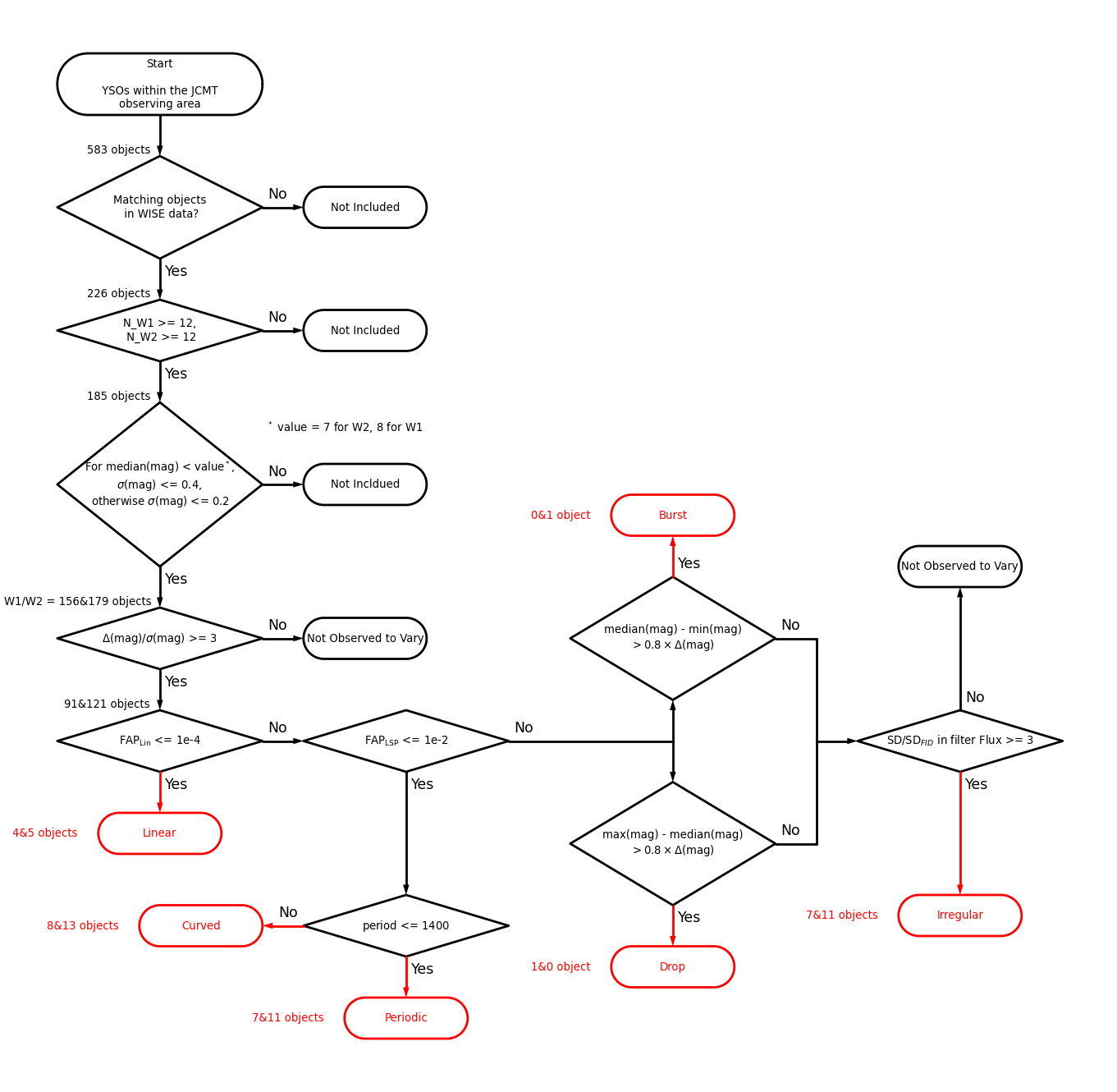}
    \caption{Summary flow chart outlining the {mid-IR} sample selection and light curve variability classification process. 
    Our methodology, based on the approach used by \citet{w.park2021}, identifies six types of variables, secular (linear, periodic, curved) and stochastic (burst, drop, irregular), among known YSO candidates in the WISE W1 and W2 bands (refer to Section~\ref{sec:analy_IR} for additional details). 
    The quantity $\mathrm{\Delta(mag)}$ represents the difference between the maximum and minimum values of the magnitude, while SD and $\mathrm{\sigma(mag)}$ (or simply $\sigma$) denote the standard deviation and mean uncertainty of magnitudes across all epochs for a given source, respectively.
    This methodology was applied to each band separately. {The number of objects that satisfy each flow chart criteria are indicated in the figure. Where two numbers are provided, the first is for W1 and the second for W2.}} 
\label{fig:IR_flowchart}
\end{figure*}

In accordance with the YSO class nomenclature outlined by \citet{w.park2021}, we classify 156 objects in W1 and 179 objects in W2 into four categories: {Class 0/I\footnote{For this classification also includes flat SED YSOs.}; (33.8\%\ in W1 and 34.4\%\ in W2), Class II (60.5\%\ in W1 and 60.6\%\ in W2), Class III (3.2\%\ in W1 and 2.8\%\ in W2), and ambiguous 
(2.5\%\ in W1 and 2.2\%\ in W2).}  This information, and comparison against the full YSO sample for \region, is shown graphically in Figure~\ref{fig:IR_YSO_sample}b.


\begin{figure*}
   \centering
   \includegraphics[width=\textwidth]{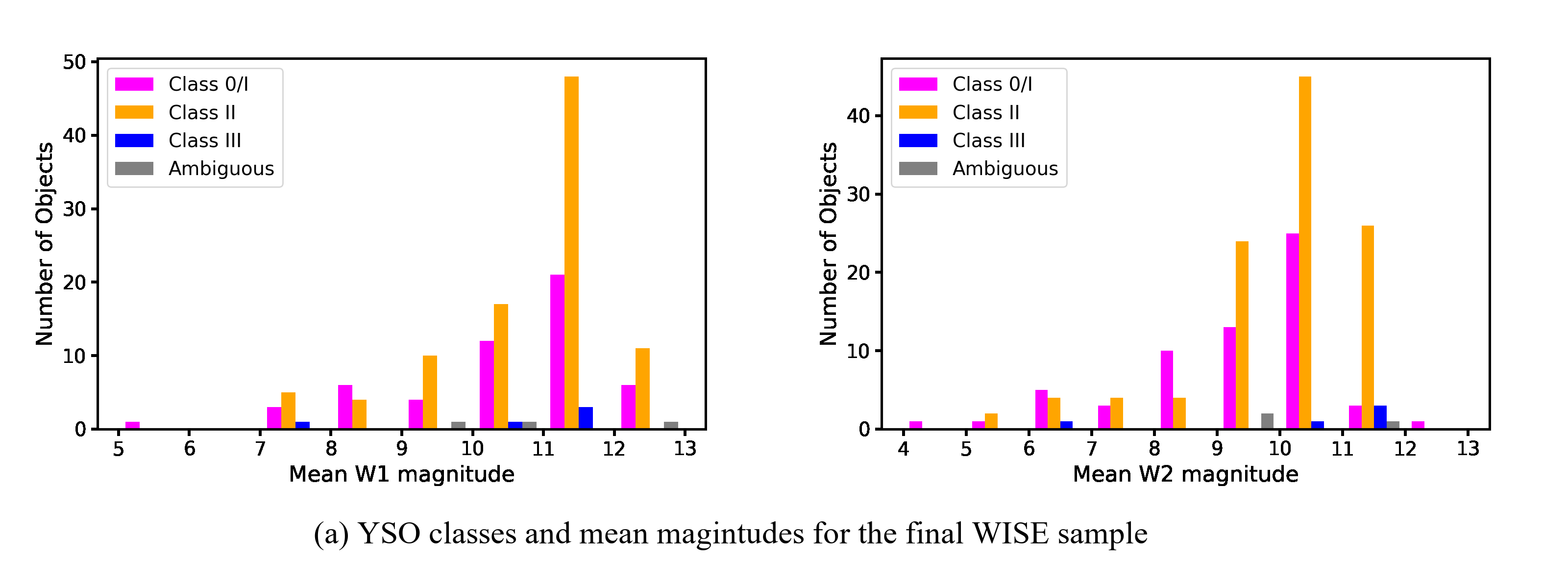}
   \includegraphics[width=\textwidth]{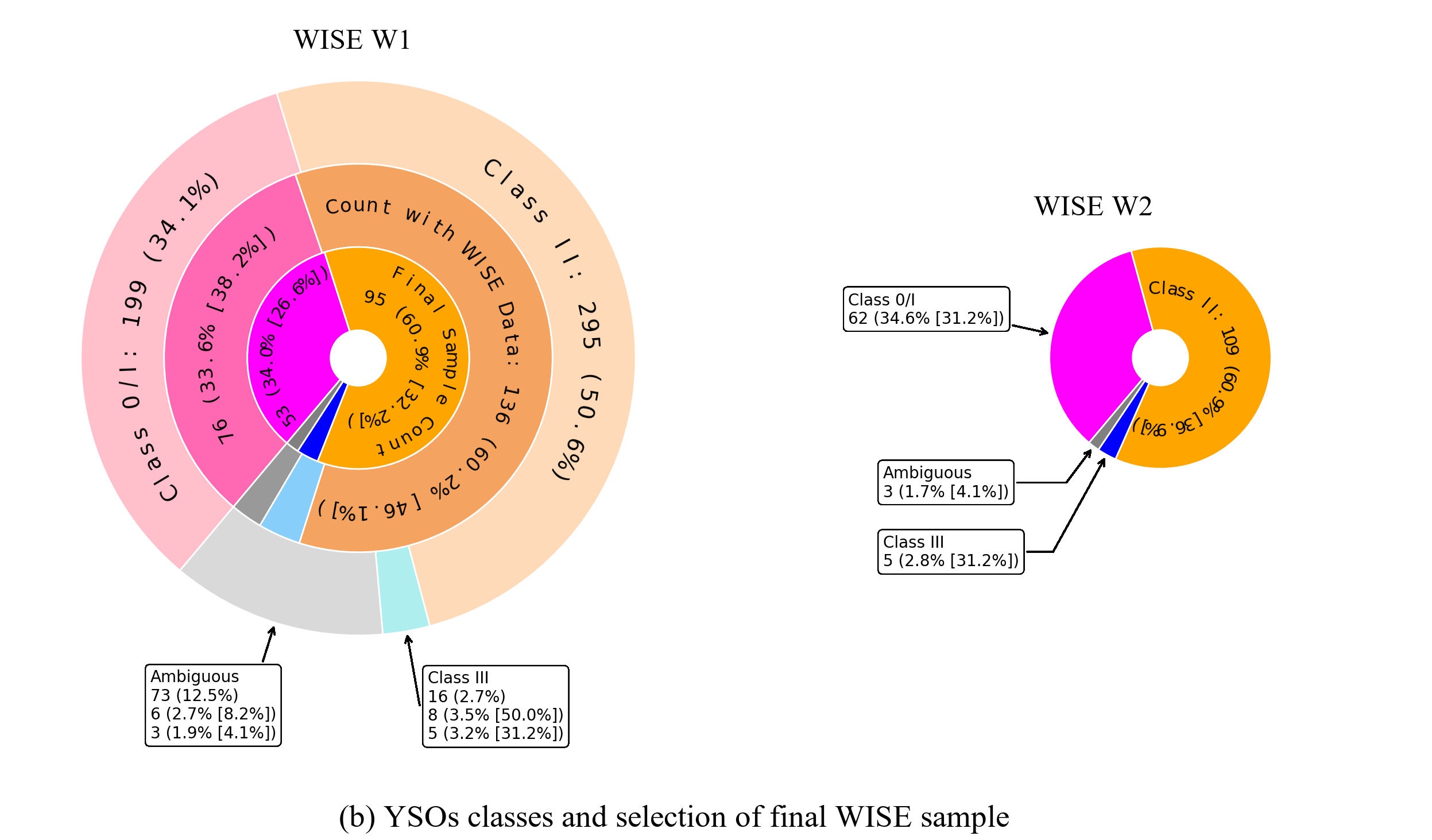}
    \caption{Overview of YSO classification and data selection: 
    (a) Distribution of mean magnitudes for the final sample of YSOs across different evolutionary stages: {Class 0/I, Class II, Class III, and ambiguous classifications}, shown in magenta, orange, blue, and gray, respectively.
    (b) Nested pie charts representing population statistics for YSOs at various stages of data collection and analysis for both WISE W1 and W2 bands.
    Each segment within the charts corresponds to a distinct YSO type, with the outer ring representing the total count of that type, the middle ring showing those with WISE data, and the innermost ring indicating the count in the final sample.
    The percentages on the pie charts reflect the proportion of each YSO type within the respective layer and the percentage of the total count for each YSO type.
    }
    \label{fig:IR_YSO_sample}
\end{figure*}

The mid-IR light curve analysis for low-mass protostars performed by \citet{w.park2021} focused on the WISE W2 band, and identified variable sources exhibiting various types of behavior. 
In this study we slightly modify this technique to fit the \region\ sample, and apply the analysis to both W1 and W2 data sets. {In the first step to analyse variability, we convert W1 and W2 magnitudes to fluxes. From these, we estimate a standard deviation (SD) and a mean flux uncertainty \citep[$\sigma$ or SD$_{fid}$, see][]{Carlos2020,w.park2021}. The value of $\mathrm{SD/SD_{fid}}$ serves as a normalized measure of variability, accounting for the expected fluctuations in the data. Previous works have used this parameter and found that selecting sources with $\mathrm{SD/SD_{fid}}>$3, provides a reliable list of candidate variable YSOs \citep{Carlos2020}.

Figure~\ref{fig:IR_SDoverSDfid} shows the amplitude of variability $\Delta$(max-min), the difference between the maximum and minimum magnitudes, versus $\mathrm{SD/SD_{fid}}$ for both W1 and W2. Similar to the conclusions by \citet{w.park2021}, the figure shows that numerous sources exhibit strong brightness variations even when their $\mathrm{SD/SD_{fid}}$ values fall below 3. 
This highlights the importance of thoroughly analyzing all sources, even those with lower $\mathrm{SD/SD_{fid}}$ values, to understand their variability behavior fully.}

\citet{w.park2021} classified variable behavior into six distinct categories, which can be grouped into three secular and three stochastic variability categories. 

\noindent Secular variability includes the following behaviors:
\begin{itemize}
    \item Linear: These sources display straight-line changes in brightness over time, either increasing or decreasing steadily;
    \item Periodic: These sources exhibit regular, repeating brightness variations over time. 
    Such periodicity can arise from a variety of astrophysical phenomena, including cycles of long-duration obscuration, binary interactions, or repetitive bursts.
    These mechanisms often involve interactions within the star-disk system or between binary components, leading to observable periodic changes in the systems' overall brightness \citep[e.g.,][]{bouvier2013, grankin2008, herbst10,cody2014, hodapp2012, yoo2017};
    \item Curved: These sources show continuous, non-linear brightness changes that do not repeat over our time baseline.  This variability may arise from evolving circumstellar material, disk instabilities, or inner disk height changes.
    Some curved objects might later be reclassified as periodic with longer observations.
\end{itemize}
\noindent Stochastic variability includes:
\begin{itemize}
    \item Burst: These sources exhibit sudden, short-lived increases in brightness, potentially resulting from episodic accretion events or flaring activity;
    \item Drop: These sources display abrupt, transient decreases in brightness that may be attributed to brief extinction events, possibly arising from the geometric properties of disks;
    \item Irregular: These sources show unpredictable, erratic variations in brightness without any discernible pattern, possibly due to a combination of several underlying mechanisms or chaotic processes. A sparse sampling of light curves may cause some YSOs with periodic variability to be misclassified as irregular \citep{y.lee2020}.
\end{itemize}

Following the schematic flowchart in Figure \ref{fig:IR_flowchart}, we first examined sources with $\mathrm{\Delta(max-min)/\sigma(mag)} \geq 3$ to investigate the presence and type of variability.
Here, $\mathrm{\Delta(max-min)}$ represents the difference between the maximum and minimum values of the W1 or W2 magnitude, and $\mathrm{\sigma(mag)}$ denotes the mean uncertainty of the W1 or W2 bands across all epochs for a given source.
The criterion of {$\mathrm{SD/SD_{fid}} > 3$} is commonly used to identify variable sources.
However, this threshold may not be effective in detecting all types of variability.
For instance, secular variables may have low standard deviation due to an underlying regular pattern and weak slope, while stochastic variability may occur at only one epoch, leading to a low standard deviation over the entire light curve.
To address these issues, \citet{w.park2021} set an alternative criterion of $\mathrm{\Delta mag/\sigma(mag)} \geq 3$ in the magnitude domain to search for variability.
Out of 156 sources at W1, we find 91 or 58\% {that should be further checked for  
variability}.  {At W2, 121 of 179 (68\%) are potential variables.}

\begin{figure}
    \centering
   \includegraphics[width=.49\textwidth]{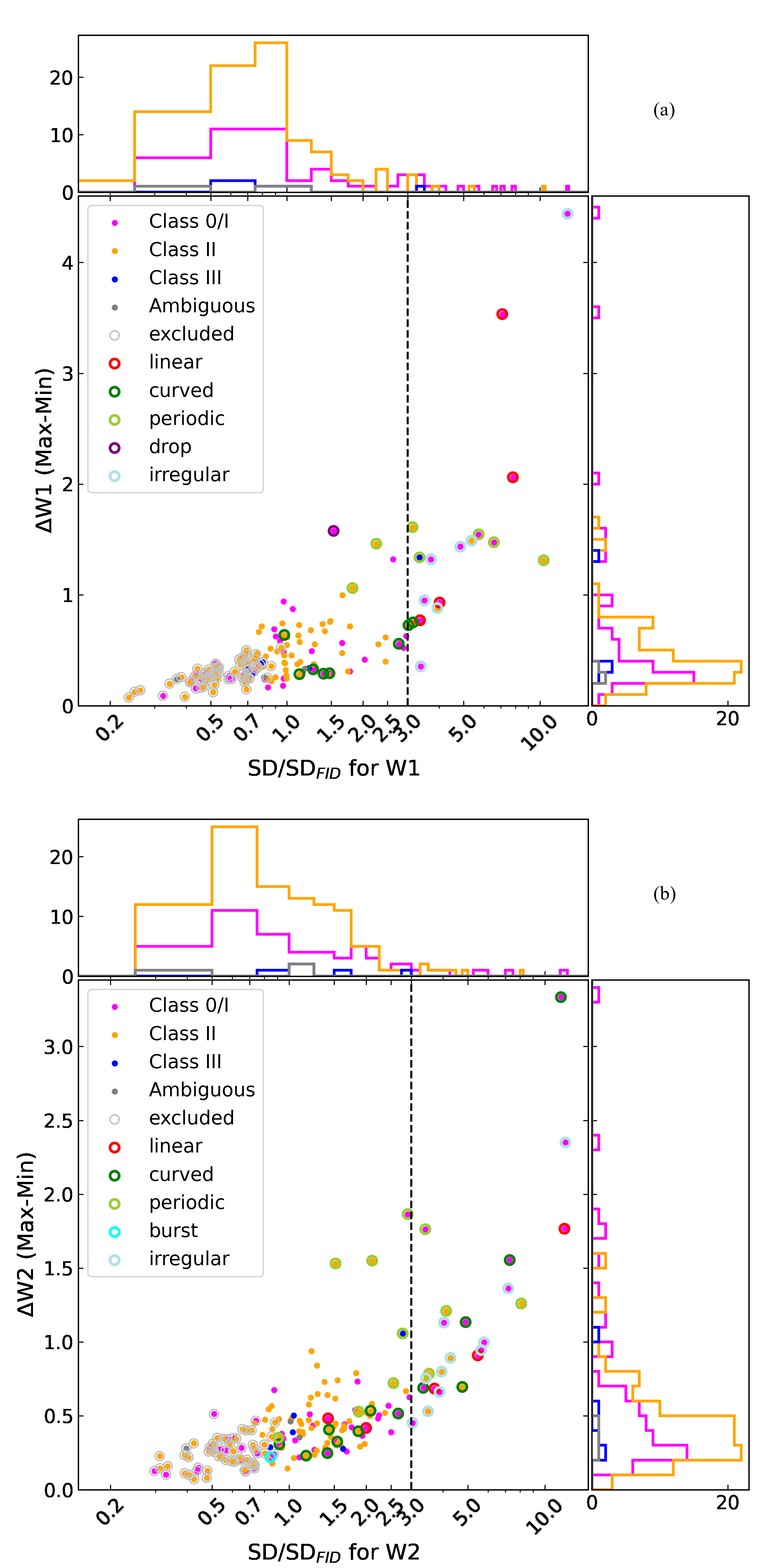}
    \caption{Comparison of $\mathrm{SD/SD_{fid}}$ and $\mathrm{\Delta mag}$ for all sources in the final samples of W1 (a) and W2 (b).
    The dashed vertical line indicates the commonly used threshold value of $\mathrm{SD/SD_{fid}}$ equal to 3 for identifying sources with brightness variability.
    The classification of YSOs is indicated by the filled circle colors, which are consistent with those used in Figure~\ref{fig:IR_YSO_sample}. 
    Variable candidates among the YSOs are denoted by a bordered circle, with the color of the border circle indicating the type of variability, while sources that were excluded during the analysis because they have a $\mathrm{\Delta mag/\sigma(mag)} < 3$ are marked with the gray bordered circle.}
    \label{fig:IR_SDoverSDfid}
\end{figure}

The Lomb-Scargle periodogram \citep[LSP;][]{lomb1976, scargle1989} and a linear fitting are the secular models used in our analysis.
The robustness of the secular variability is evaluated from the false alarm probability (FAP).
The LSP FAP quantifies the likelihood of identifying a spurious signal in an LSP, whereas the linear FAP estimates the probability of falsely detecting a linear trend. 
However, the traditional LSP FAP can provide a systematic overestimate of the false alarm probability for long periods \citep{w.park2021,y.lee2021}. 
Therefore for the LSP fit we used a modified version, which though labelled $\mathrm{FAP_{LSP}}$ takes that same form as the FAP$_{\rm mod}$ introduced by \citet{y.lee2021}.

In Figure~\ref{fig:IR_FAP}, we compare the FAP for linear and periodic fits for our \region\ samples at W1 and W2.
Sources with $\mathrm{FAP_{Lin}} \leq$ $10^{-4}$ are classified as exhibiting linear variability.
Among the remaining sources, those with $\mathrm{FAP_{LSP}} \leq$ $10^{-2}$ and a period of 1400 days or shorter, which corresponds to about half of the eight years covered by the WISE data used in our analysis, are identified as displaying periodic variability.
Conversely, sources with $\mathrm{FAP_{LSP}} \leq$ $10^{-2}$ and a period longer than 1400 days are classified as exhibiting curved variability.
From the W1 sample, we identify four linear, nine curved, and seven robust periodic variables.
The W2 sample yields five linear, thirteen curved, and eleven periodic variables.

\begin{figure}
    \centering
   \includegraphics[width=.5\textwidth]{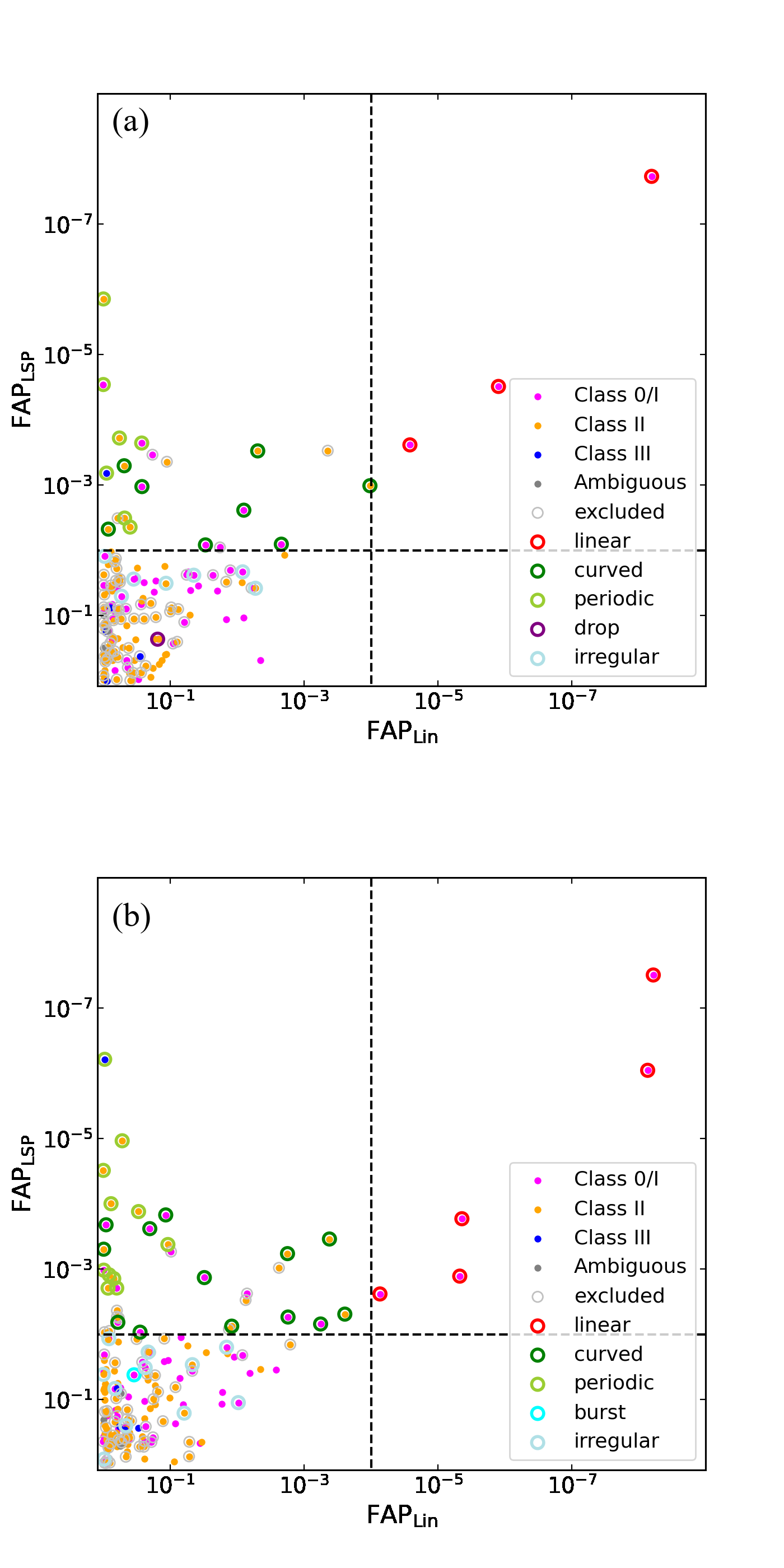}
    \caption{Comparison of the FAPs for linear and periodic fits for all sources in the final samples of W1 (a) and W2 (b).
    The dashed vertical line represents a FAP of $10^{-4}$ for the linear fit, while the dashed horizontal line represents a FAP of $10^{-2}$ for the periodic fit. 
    Symbols and colors used in this figure are consistent with those presented in Figure~\ref{fig:IR_SDoverSDfid}, representing different evolutionary stages and the five types of mid-IR flux variability identified in our study.}
    \label{fig:IR_FAP}
\end{figure}

We apply an additional constraint to identify sources that exhibit either bursts or drops in brightness while maintaining stable fluxes over the remaining epochs.
Specifically, we require either $\mathrm {median(mag) - min(mag)}$ $> 0.8 \times \mathrm{\Delta(mag)}$ for burst sources or $\mathrm{max(mag) - median(mag)}$ $> 0.8 \times \mathrm{\Delta(mag)}$ for drop sources.
After excluding all previously identified sources, those with $\mathrm{SD/SD_{fid}}$ $\geq 3$ are classified as showing irregular variability.

For the W1 sample, we identified one drop and seven irregular variables but no burst variable.
The W2 sample, on the other hand, yielded one burst and eleven irregular variables but no drop variable.
Further discussion and analysis of these findings, supported by light curve data, is presented in Section~\ref{sec:disc_IR}.

Table~\ref{tab:variables_mir_type} summarizes the statistical results from our independent variability investigations using WISE W1 and W2.
We present the YSOs that are robust mid-IR variables across the six types, along with their corresponding evolutionary stages.
Combined, 47 YSOs exhibit robust variability at either W1 or W2, including 21 YSOs that vary in both bands. 
Interestingly, while 14 display the same type of variability in both bands, 7 are classified into somewhat different variability types.
Upon closer examination of the light curves for these seven cases, multiple phenomena appear to be at play.
The remaining 26 YSOs are variable in just one of the two bands, with 7 variable in W1 and 20 in W2.
This disparity in variability across the two bands will be examined in further detail, including an analysis of their light curves, in Section~\ref{sec:corr_w1nw2}.
For completeness, we present in  Table~\ref{tab:variables_mir_sources} and \ref{tab:nonvariables_mir_sources}, the variability measures derived from the NEOWISE light curves for both the final sample of variable candidates identified in this study (Table~\ref{tab:variables_mir_sources}) and the remaining sources from the final sample that were not observed to vary (Table~\ref{tab:nonvariables_mir_sources}).

\begin{deluxetable*}{lrlrl rlrlc crlrl rlrlc}
\tabletypesize{\scriptsize}
\setlength{\tabcolsep}{1mm}
\tablecaption{Variable Type by YSO Classification in WISE W1 and W2 \label{tab:variables_mir_type}}
\tablehead{
 & \multicolumn{9}{c}{W1} & & \multicolumn{9}{c}{W2} \\
{Type} & \multicolumn{2}{c}{(Class 0/I, FS)} & \multicolumn{2}{l}{(Class II)} & \multicolumn{2}{l}{(Class III)} & \multicolumn{2}{l}{(ambiguous)} & \colhead{Total} &&
\multicolumn{2}{l}{(Class 0/I, FS)} & \multicolumn{2}{l}{(Class II)} & \multicolumn{2}{l}{(Class III)} & \multicolumn{2}{l}{(ambiguous)} & \colhead{Total} \\
\cline{1-20}
{Count} & \multicolumn{2}{l}{53} & \multicolumn{2}{l}{95} & \multicolumn{2}{l}{5} & \multicolumn{2}{l}{3} & \colhead{156} &&
\multicolumn{2}{l}{62} & \multicolumn{2}{l}{109} & \multicolumn{2}{l}{5} & \multicolumn{2}{l}{3} & \colhead{179} 
}
\startdata
Linear                                  &  4 & (7.5)  &  0 & (0)    & 0 & (0)  & 0 & (0)  &   4     &&  5 & (8.1)  &  0     & (0)    & 0 & (0)    & 0 & (0)    &   5 \\
Curved                                  &  4 & (7.5)  &  4 & (4.2)  & 0 & (0)  & 0 & (0)  &   8     &&  8 & (12.9) &  5     & (4.6)  & 0 & (0)    & 0 & (0)    &  13\phn \\
Periodic                                &  2 & (3.8)  &  4 & (4.2)  & 1 & (20) & 0 & (0)  &   7     &&  2 & (3.2)  &  8     & (7.3)  & 1 & (20)   & 0 & (0)    &  11\phn \\
Burst                                   &  0 & (0)    &  0 & (0)    & 0 & (0)  & 0 & (0)  &   0     &&  1 & (1.6)  &  0     & (0)    & 0 & (0)    & 0 & (0)    &   1 \\
Drop                                    &  1 & (1.9)  &  0 & (0)    & 0 & (0)  & 0 & (0)  &   1     &&  0 & (0)    &  0     & (0)    & 0 & (0)    & 0 & (0)    &   0 \\
Irregular\tablenotemark{\scriptsize a}  &  5 & (9.4)  &  2 & (2.1)  & 0 & (0)  & 0 & (0)  &   7     &&  7 & (11.3) &  4     & (3.7)  & 0 & (0)    & 0 & (0)    &  11\phn \\
\hline                                                                                       
All Variables                                   & 16 & (30.2) & 10 & (10.5) & 1 & (20) & 0 & (0)  &  27\phn && 23 & (37.1) &  17    & (15.6) & 1 & (20)   & 0 & (0)    &  41\phn \\
\enddata
\tablecomments{Numbers are the count of variables for each variable type, while numbers in parentheses are the fraction (\%) of variable candidates relative to the numbers in the ``Count" row, which denote the selected WISE samples in each evolutionary stage (see Section~\ref{sec:analy_IR}).}
\tablenotetext{a}{Sources with $\mathrm{SD/SD_{fid}}$ $\geq 3$ but not classified as any previous specific type of variability.}
\end{deluxetable*}

\section{Discussion} \label{sec:disc}

\subsection{Spatial Coincidence of Sub-mm and Mid-IR Sources} \label{sec:source_loc}

This study focuses on the variability of sources at both sub-mm and mid-IR wavelengths within the same field of view centered on \region. While the two wavelengths may respond to different time-dependent processes, the expectation is that for YSOs the long-term light curves are indicators of any underlying mass accretion variability onto the central source. 
Previously, such connections between mid-IR and sub-mm have been observed for nearby star-forming protostars by \citet{Carlos2020,y.lee2020,Yoon2022}.

Figure~\ref{fig:map_vari_loc} illustrates the spatial distribution of source peaks in our observed area at both sub-mm and mid-IR wavelengths. 
The co-added 850~\um\ map from 23 epochs in Figure~\ref{fig:map_vari_loc}a reveals the sub-mm peak locations and the three robust sub-mm variables.
Similarly, Figures~\ref{fig:map_vari_loc}b--\ref{fig:map_vari_loc}c display the WISE W1 and W2 images from the first epoch (March 2014), showing the locations of all YSO candidates located within our JCMT surveyed area and highlighting the mid-IR variables.
The sub-mm sources with a mean peak flux exceeding 100~$\mjybeam$ usually reside within sub-mm filamentary structures.
Furthermore, numerous WISE-associated YSOs are situated along these same filaments, as expected for stars forming out of the prenatal molecular cloud material.
Notably, our observation area also includes some isolated YSOs identified by \citet{kuhn2021} that may not have a direct association with \region.

While we identify variability independently for each wavelength, we specifically examine the spatial correlation of the sources without delving into their detailed properties.
We chose a maximum separation of 3$^{\prime\prime}$ between the sub-mm peaks and the YSO candidates when matching. 
This value is based on the relative pointing accuracy of the JCMT, which is typically better than 2$^{\prime\prime}$ \citep{Mairs2024}, the localized nature of the mid-IR fluxes, and the desire for the peaks to remain within a typical core scale (5000 au) separation.
Only seven YSOs from the \citet{povich2010} and \citet{kuhn2021} catalogs match the location of sub-mm peaks in the SCUBA-2 images.
Among these seven matched YSOs, only three have corresponding WISE measurements.
Finally, among the three matching YSOs, no source shows sub-mm flux variation, even though two of the sources, SPICY 79425 and 80368, 
show mid-IR flux variability in both WISE bands.
This matching compares the observed sub-mm peaks with YSOs that also have NEOWISE measurements within a tight matching distance of $\sim$5000\,au.
YSOs that reside within the sub-mm filaments may not live at the peaks, since this may be where new star formation is about to occur. 
Similarly, sub-mm peaks may be optically thick to mid-IR wavelengths and thus may hide the most recent star formation. 
Both of these situations are seen in nearby star-forming regions, where angular resolution allows for a clearer investigation of the connection between the sub-mm and mid-IR images.

Some additional difference between the sub-mm and mid-IR variability may be introduced by the distinct observation intervals and total monitoring periods.
These two datasets only overlap during the last three years of the WISE observations. 
Furthermore, the cadence in the sub-mm is roughly monthly whereas in the mid-IR the epochs are only twice per year.
Thus, the sub-mm has a much higher temporal resolution than the mid-IR, while the mid-IR has an extended observational baseline of about nine years, three times longer than the sub-mm observations considered here.

%
%
\begin{figure*}
    \centering
    \includegraphics[width=.49\textwidth]{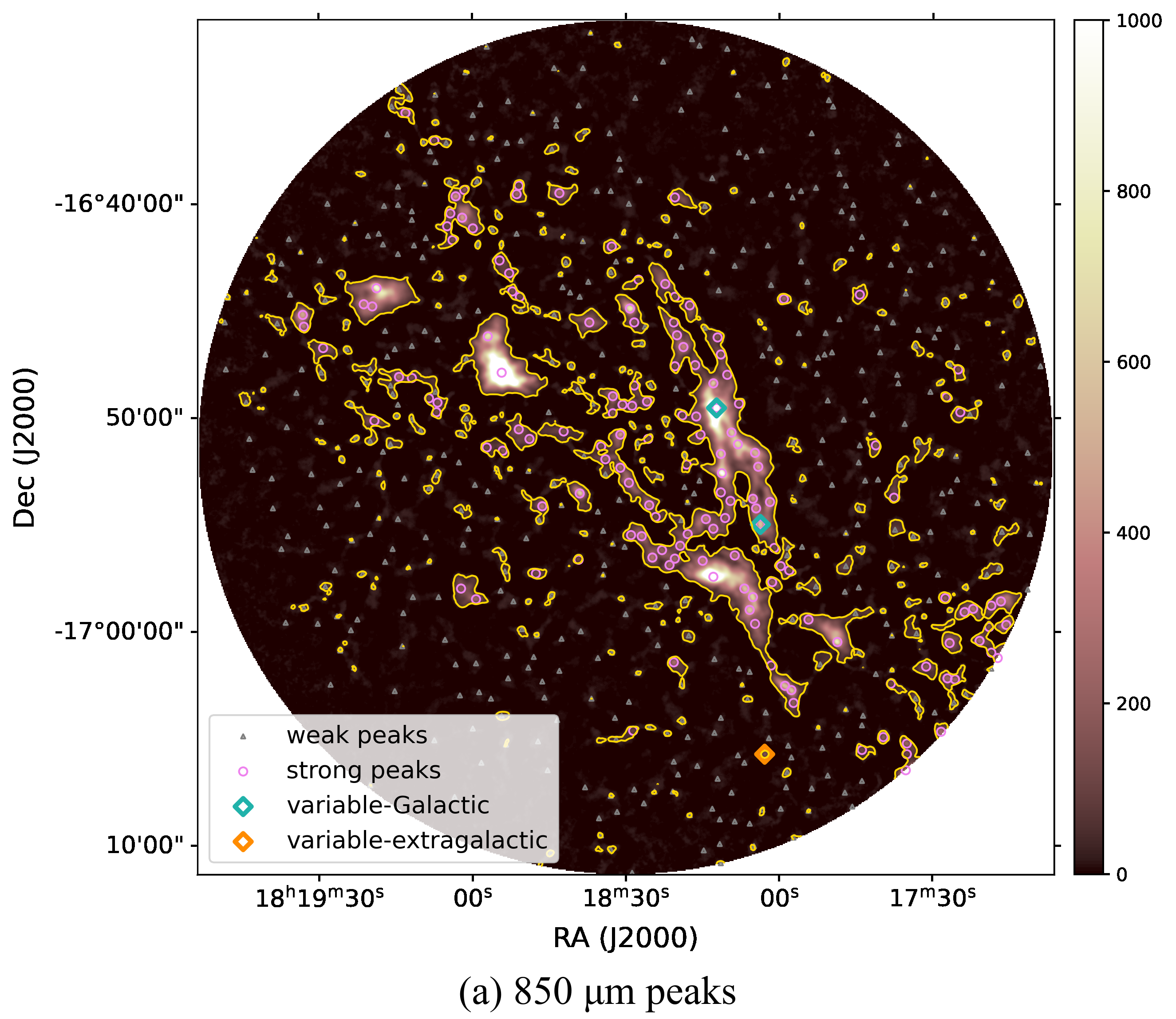}
    \begin{minipage}{\textwidth}
        \centering
        \includegraphics[width=.49\linewidth]{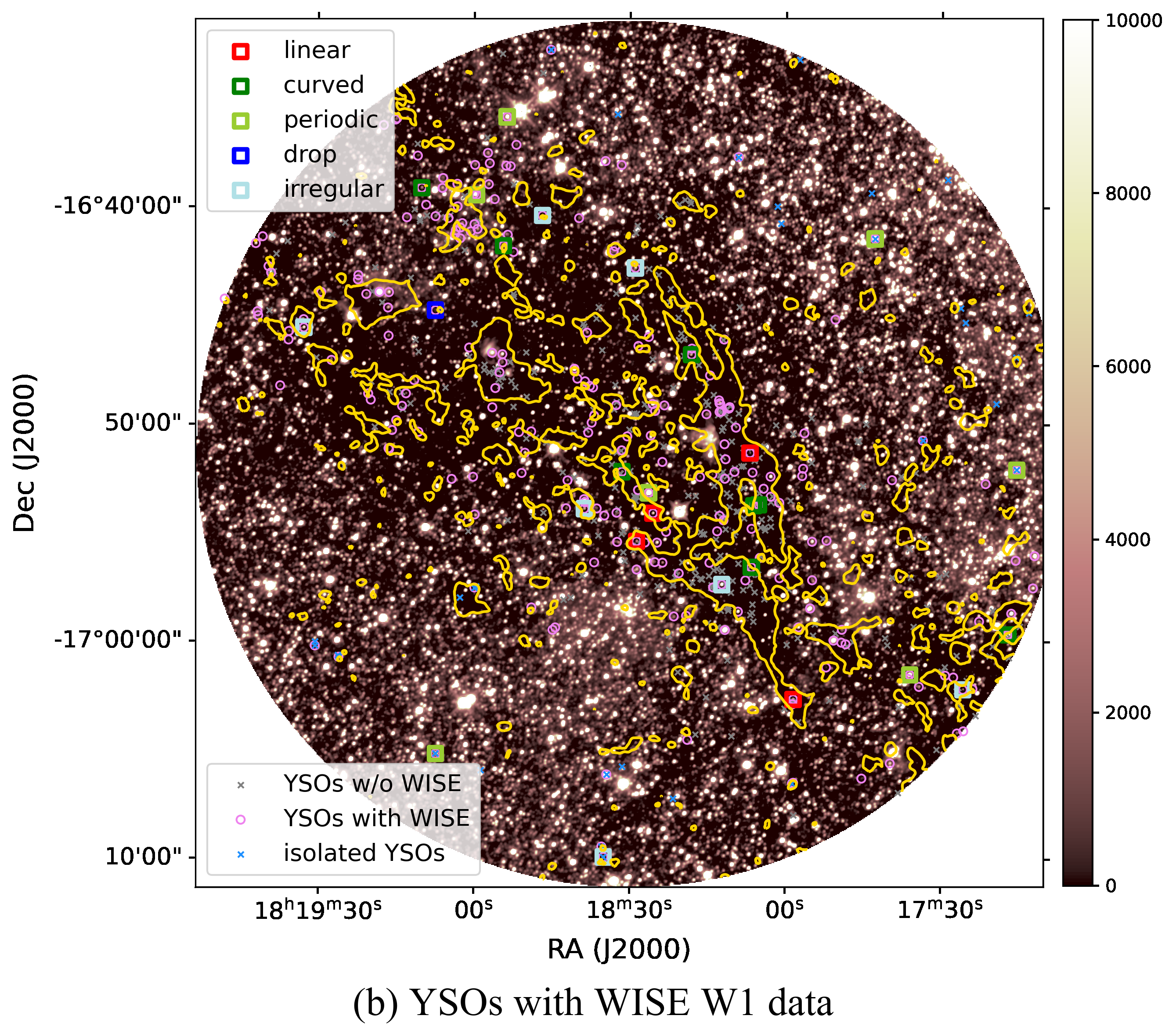}
        \includegraphics[width=.49\linewidth]{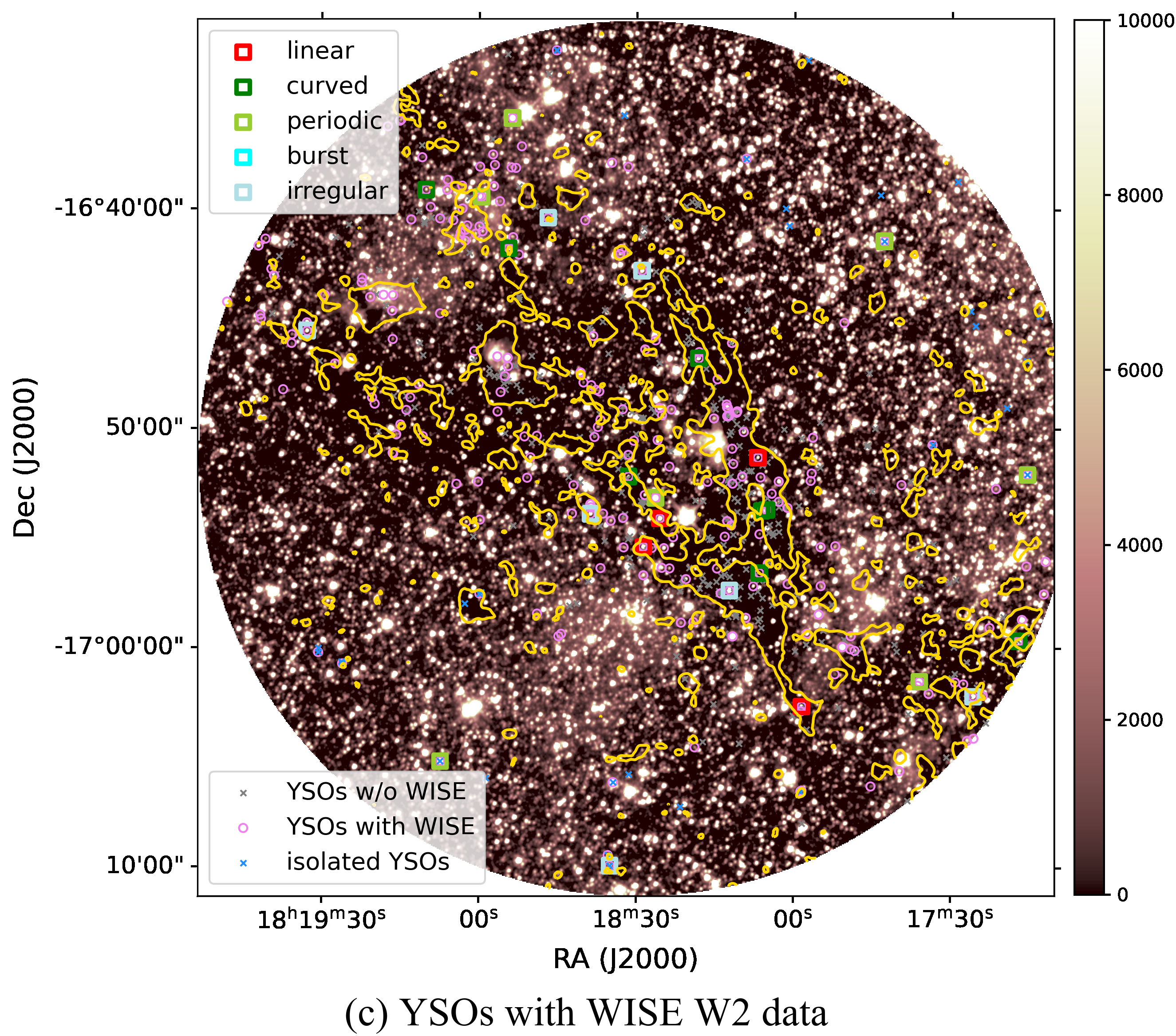}
    \end{minipage}
    \caption{
    Positions of target source peaks within the \region\ observation area at sub-mm (a) and mid-IR (b--c).
    The 850~\um\ image in (a) is a co-added image from 23 epochs, with a 30~$\mjybeam$ contour overlaid. 
    Target source peaks with a mean peak flux greater than 100~$\mjybeam$ are marked with {pink circles}, while those with a mean peak flux less than  100~$\mjybeam$ are shown with gray triangles.
    Three sub-mm robust variables are highlighted with {teal and orange diamonds}.
    The mid-IR images in (b--c) show WISE W1 and W2 data acquired during the first epoch used (March 2014), overlaid with the 850~\um\ contour shown from the (a) image.
    YSOs within our JCMT-observed area are marked: {pink} circles indicate those identified with WISE data, while gray crosses indicate those not identified. 
    Additionally, royal-blue crosses mark YSOs cataloged by \citet{kuhn2021} that were not included in any group in their study.
    Colored squares highlight mid-IR variables identified in this study.
    The color scales are displayed on the right side of each map, and the units are (a) $\mjybeam$ and (b,c) arbitrary counts.}
    \label{fig:map_vari_loc}
\end{figure*}

\subsection{Individual Sub-mm Variables} \label{sec:disc_submm}

In our study of the \region\ region, three sources exhibiting sub-mm variability at 850~\um\ are identified (see Table \ref{tab:variables_submm}). 
For simplicity, we here refer to these sources by just their right ascension, for example JCMTPP\_J181812.2.
To investigate the connection between the observed sub-mm flux variability and previously identified star-forming activity, we searched for nearby YSOs within a radius of 6\arcsec\ of the three sub-mm variable candidates.

The first source, JCMTPP\_J181812.2, has a mean peak flux of $\sim 5400~\mjybeam$ with a linear decrease since 2021 (Figure \ref{fig:lc_submm}).
This bright sub-mm source is located near a YSO identified by \citet{kuhn2021}, at a separation of about 4\farcs9.
The YSO is classified with a flat SED, placing it in a transitional phase between embedded (Class I) and optically visible (Class II)\footnote{\citet{kuhn2021} uses the [4.5]-[8.0] colour of the source to determine a value of the spectral index, $\alpha$, that yields a flat SED classification. However, using the MIPS 24 $\mu$m detection and equation 7 in \citet{kuhn2021}, we determine $\alpha$=0.43, consistent with a Class I YSO. The source is too faint for a clear detection with WISE.}.
Unfortunately, this YSO is not detected in the WISE database, precluding the analysis of its mid-IR variability.
{The proximity of the YSO to our sub-mm source, despite the somewhat significant separation, raises the possibility of a physical association. Given the distance to \region\ it is likely that the sub-mm emission is tracing a larger concentration of dust and gas than a single protostellar core and therefore the peak dust emission may well be offset from any specific embedded protostar, responsible for localized heating. We thus consider JCMTPP\_181812.2 a clump housing a candidate variable protostar.}


The second sub-mm variable, JCMTPP\_J181803.7, is comparatively weaker, with a mean peak flux of about $590~\mjybeam$, and exhibits a linear increase in flux since 2020 (Figure \ref{fig:lc_submm}). Is is also undetected by WISE. 
Unlike the brighter source, this weaker sub-mm source does not have a confirmed nearby YSO {in the \citet{povich2010} or \citet{kuhn2021} catalogues}.
{The source peak is, however, within 2\arcsec\ of the ATLASGAL clump AGAL014.131-00.522\_S \citep{Schuller+2009, Contreras+2013, Urquhart+2014}, catalogued as part of an unbiased survey of the Galaxy designed to identify high-mass clumps. The source has subsequently been classified as a protostellar clump, labelled AG0821, by \citet{rathborne2016} based on association with either 4.5 $\mu$m emission or compact 24 $\mu$m emission. We therefore suggest that the sub-mm variability arises from an unseen protostellar source.}  

The third sub-mm variable, JCMTPP\_J181802.8, has the faintest peak brightness, $\sim 150~\mjybeam$, among the three and lacks any known nearby YSO. 
Instead, it is coincident, separation of 4\arcsec,  with a nearby compact radio source, NVSS J181802-170542 (also known as IVS~B1815-171 and LQA~274-017\_001), potentially of extra-galactic origin.
Long-timescale stochastic variability of this source is evident from the standard deviation analysis and a visual inspection of its light curve (see Figure~\ref{fig:lc_submm}), which shows fluctuations but no clear secular pattern.
The brightness remains stable for the first seven epochs before experiencing a significant decrease from the 7th to the 11th epoch, corresponding to about 153 days. Subsequently, the flux experiences a gradual increase accompanied by fluctuations. The light curve again decreases during the final three epochs. 
While notably variable at 850~\um, this sub-mm source is too faint to be detected in our JCMT 450~\um\ observations and lacks a visible counterpart in the WISE source catalog.
It is probable that this radio source is a quasar. Quasars are known to often show strong variability across a large range of wavelengths. 

Intriguingly, the identification of JCMTP\_J181802.8, a potential extragalactic source showing significant sub-mm variability, echoes findings from a previous study within the JCMT Transient Survey which also uncovered an extragalactic variable source, most likely a blazar \citep[Source 2864 in NGC~2023,][]{johnstone2022}, showcasing the survey's unexpected capability to detect not only YSOs but also distant active galactic nuclei (AGNs).
These findings underscore the importance of considering extragalactic contamination when interpreting variability in sub-mm observations of star-forming regions and highlights the survey's potential as a tool for identifying and studying blazars and other variable AGNs, perhaps in combination with interferometric monitoring as analysed by \citet{Bonato2018}.  


\subsection{Mid-IR Variables} \label{sec:disc_IR}

Using the classification of sources in \region\ by \citet{povich2010,kuhn2021} we find that the  {Class 0/I (protostars)} evolutionary stage of YSOs are observed to be the most variable at mid-IR wavelengths.
First, the fractional number of variable protostars is higher than for the 
{Class~II} sources, as shown in 
Table~\ref{tab:variables_mir_type}.
In the W1 band, approximately 30\% of the protostars show variability, compared to about 11\% of those at the {Class II} stages.
Similarly, in the W2 band, around 37\% of protostars are variable, while nearly 16\% of {Class II} sources exhibit variability.
Furthermore, the protostars also have the largest amplitude of variability (Figure~\ref{fig:IR_SDoverSDfid}).

As recognized by \citet{w.park2021}, many of the periodic variables might be contaminant AGB stars, especially for sources classified as {Class II or Class III stage}. 
Consideration of Table~\ref{tab:variables_mir_type} reveals that accounting for this potential contamination only strengthens the argument that earlier evolutionary stages of star formation are more likely to show mid-IR variability.
Accurate determination of genuine PMS variability and potential AGB star contamination requires a multifaceted approach, such as searching for maser emission \citep{j.lee2021} or utilizing a multi-wavelength approach \citep{groenewegen22}.

\subsubsection{Comparing Variability Across WISE Bands 
} \label{sec:corr_w1nw2}


Our analysis has identified 26 instances of variability {that is robustly detected} in only one of the WISE bands (W1 or W2), which are closely spaced in wavelength. Furthermore, of the 21 sources determined to robustly vary in both bands, 14 are assigned different variability types {in the the two bands} when using the automated classification procedure. Together this raises intriguing questions about the underlying causes of these observed differences.

Figure~\ref{fig:lc_mir_examples} shows several representative light curves, {including some} that effectively illustrate this disparity in variability between bands.
SPICY 80368 and 80463 exhibit nearly identical light curves with only subtle variations, and flux variability was confirmed in both W1 and W2 bands for these sources.
SPICY 80128 and 81009 display flux variability in both bands, but a robust detection was only achieved for W2, where the measurement uncertainties were smaller. Similarly, weak but robust variability is observed only in W2 for SPICY 81226 and GLMA G014.2370$-$00.5063.
Finally, an interesting source, SPICY 81451, shows widely divergent trends in the two WISE bands.

\begin{figure*}
    \centering
    \begin{minipage}{\textwidth}
        \centering
        \includegraphics[width=.33\linewidth]{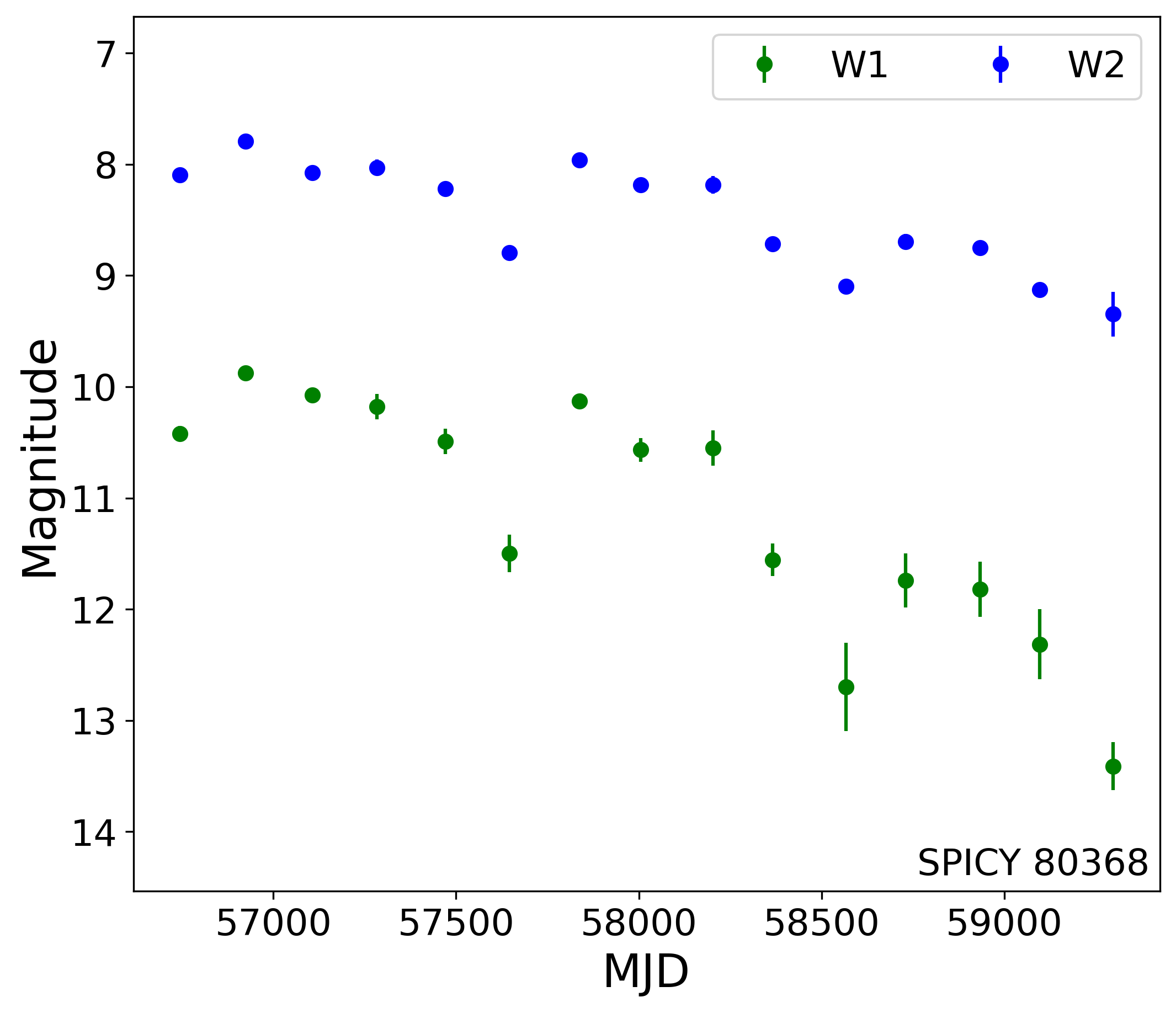}
        \includegraphics[width=.33\linewidth]{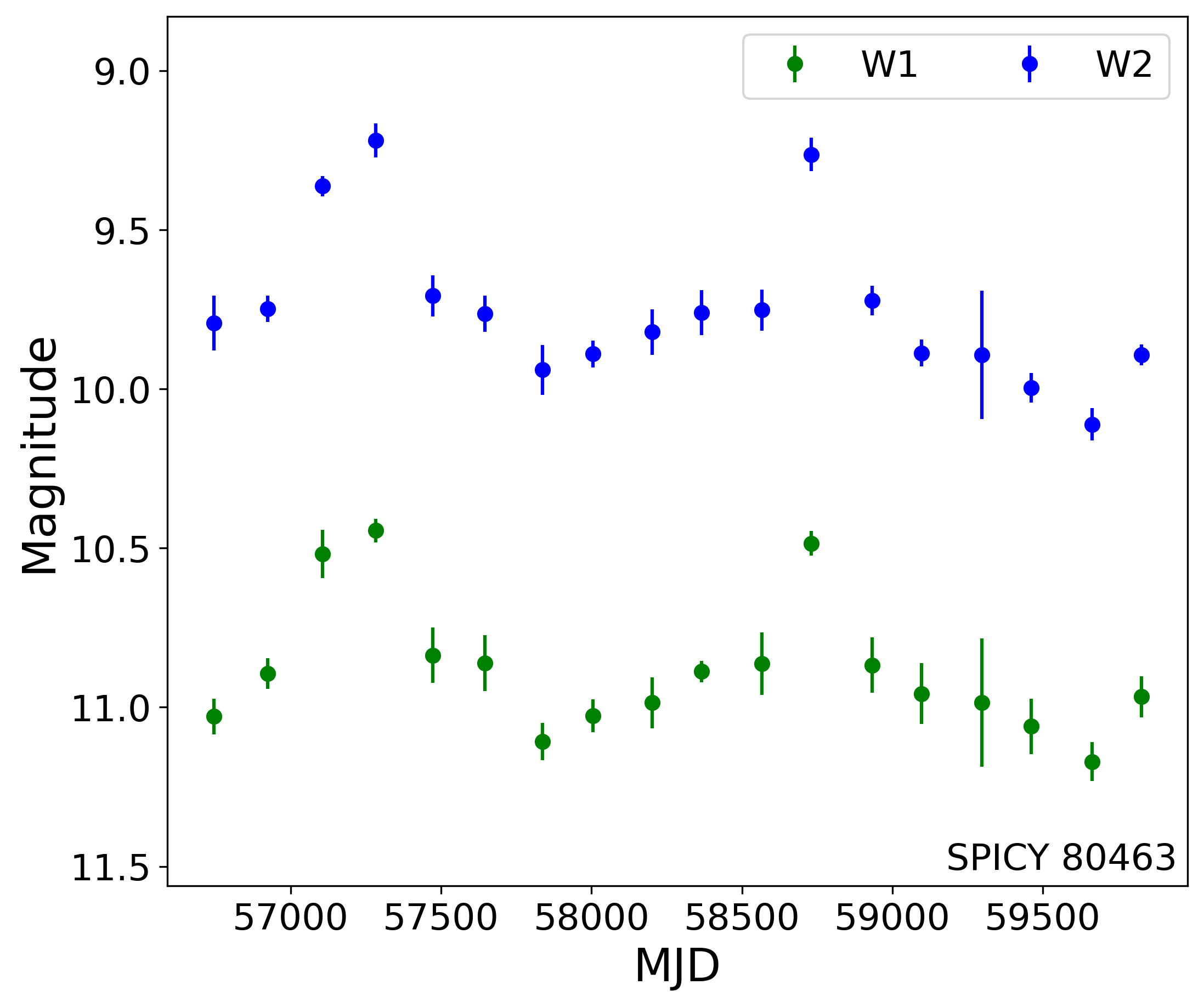}
    \end{minipage}
    \begin{minipage}{\textwidth}
        \centering
        \includegraphics[width=.33\linewidth]{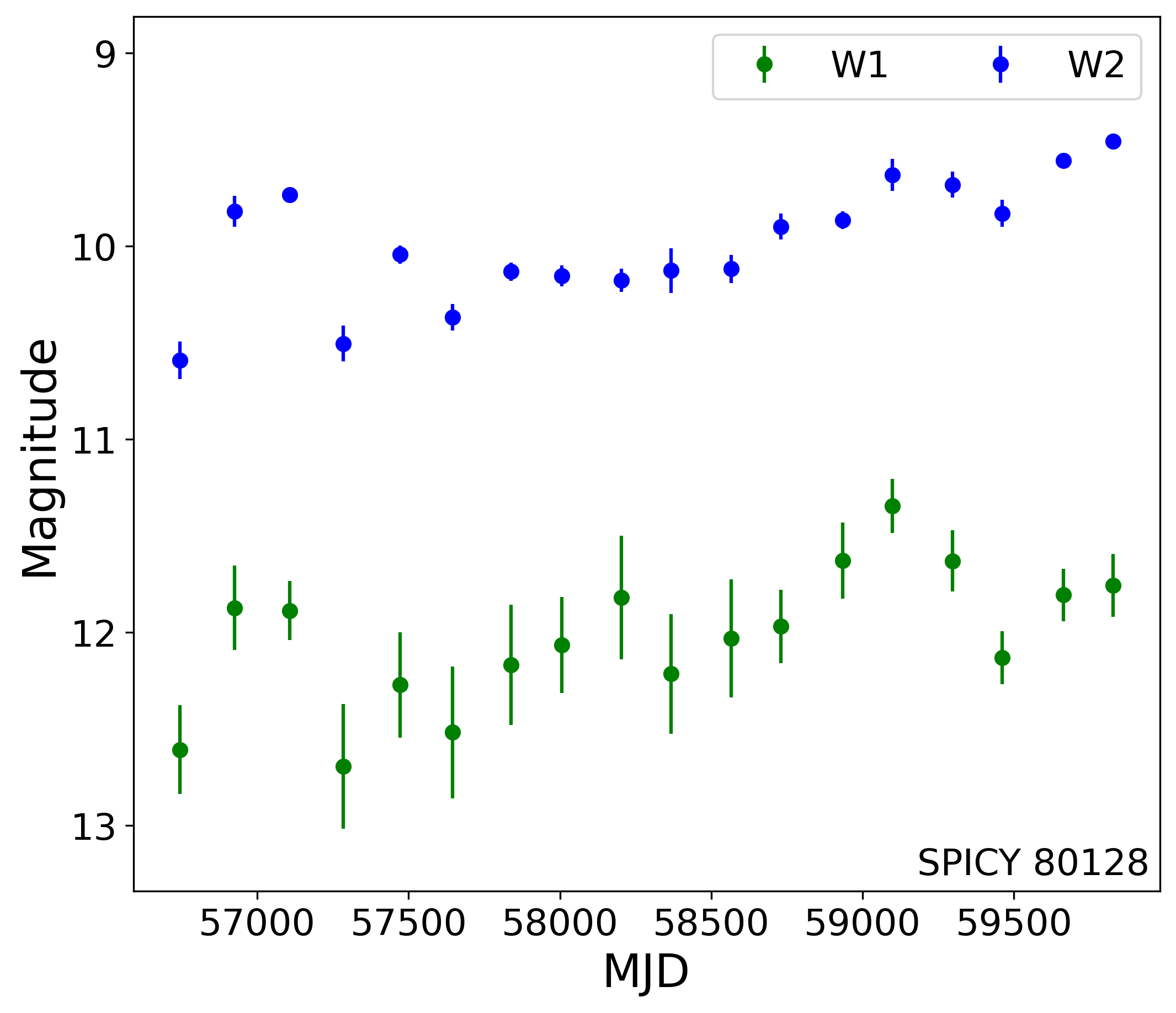}
        \includegraphics[width=.33\linewidth]{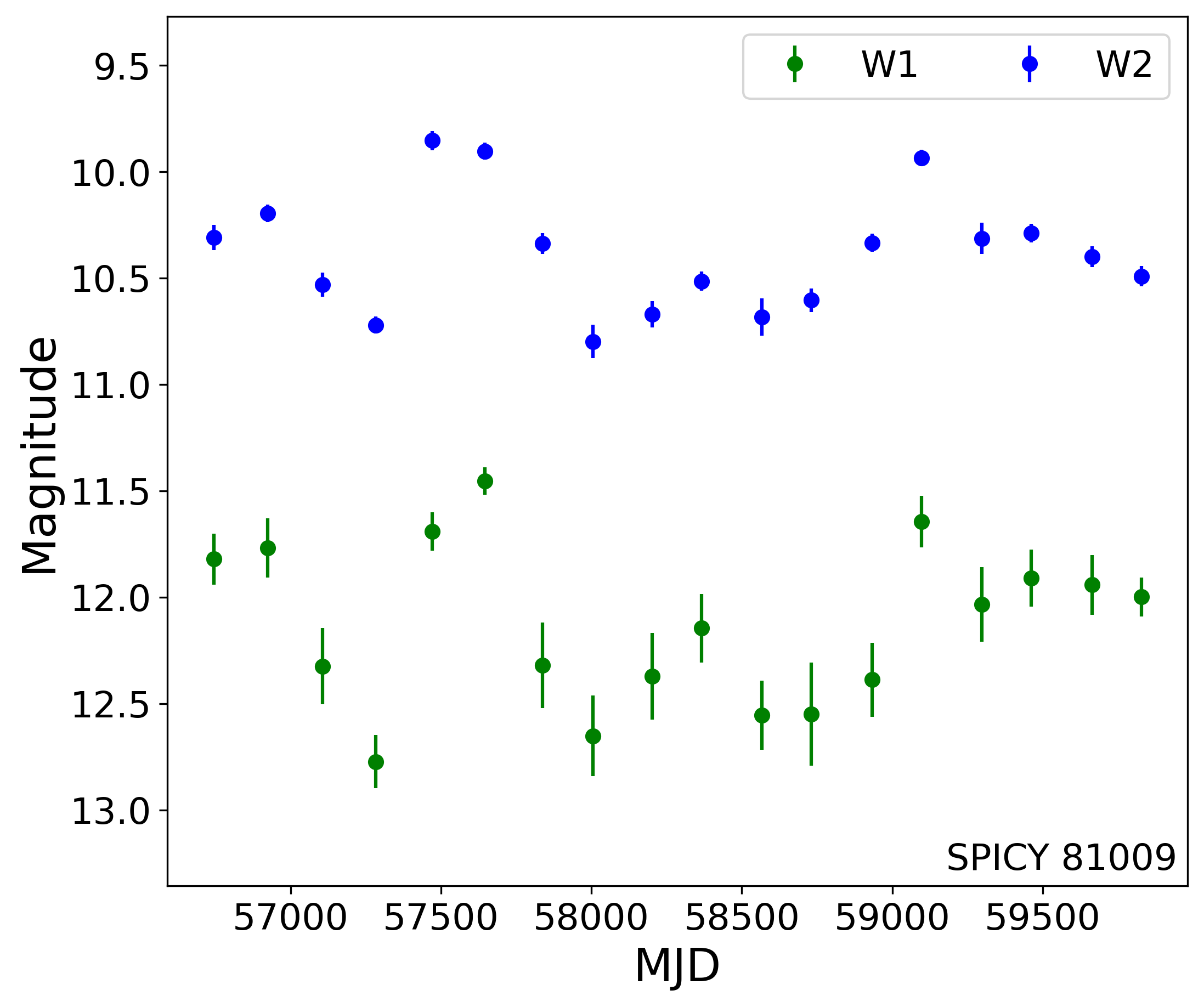}
    \end{minipage}
    \begin{minipage}{\textwidth}
    \includegraphics[width=.33\textwidth]{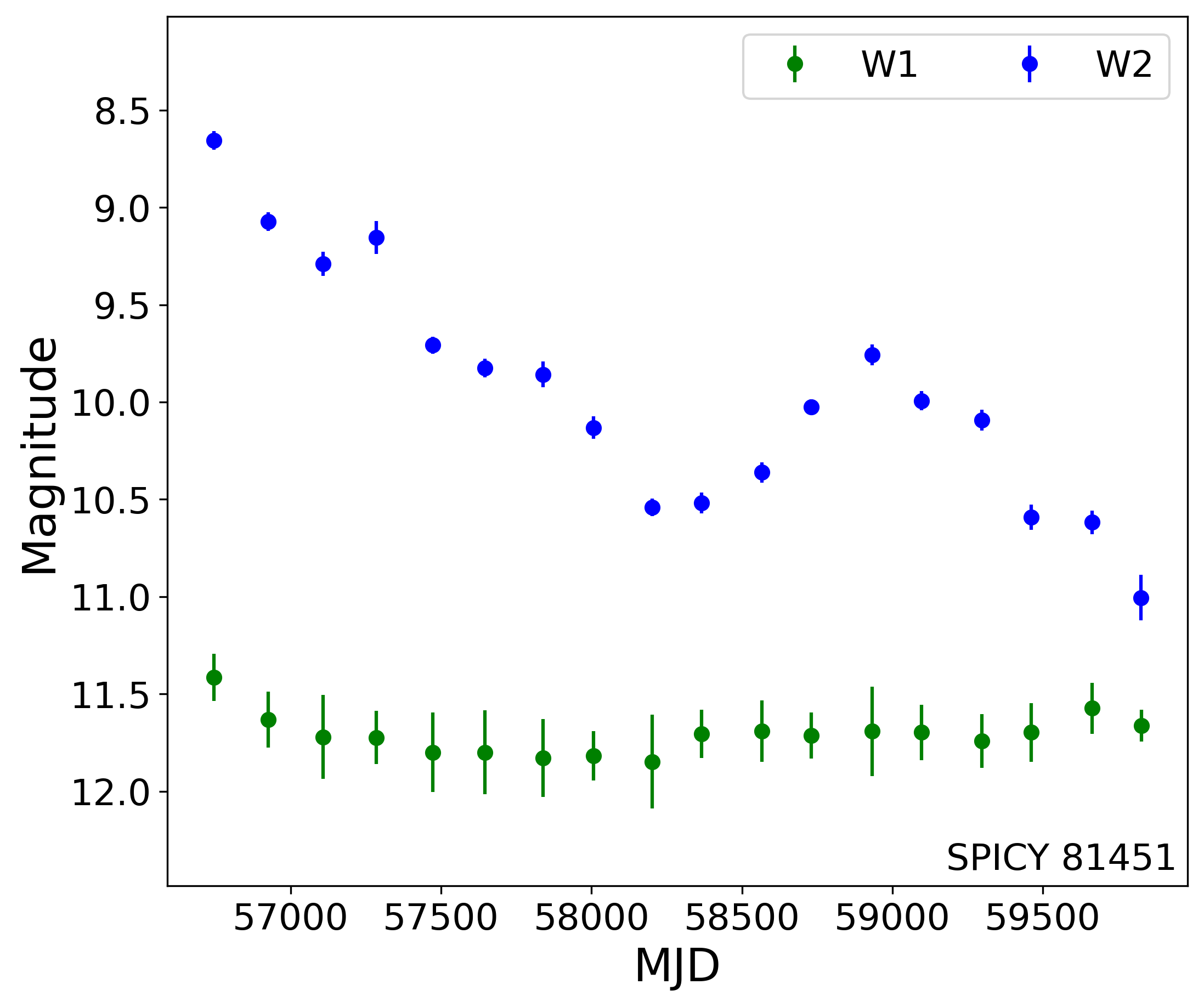}
    \includegraphics[width=.33\textwidth]{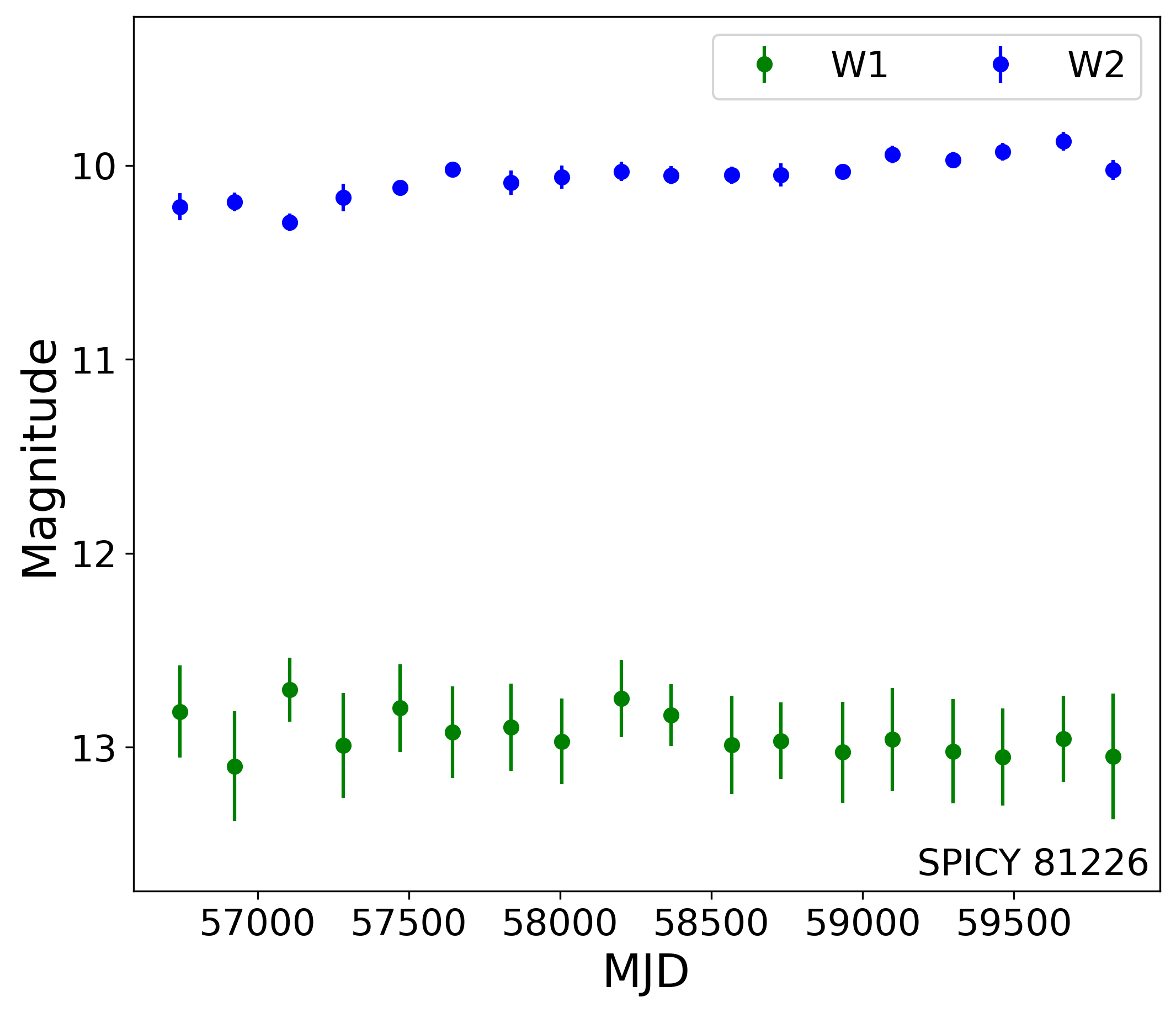}
    \includegraphics[width=.33\textwidth]{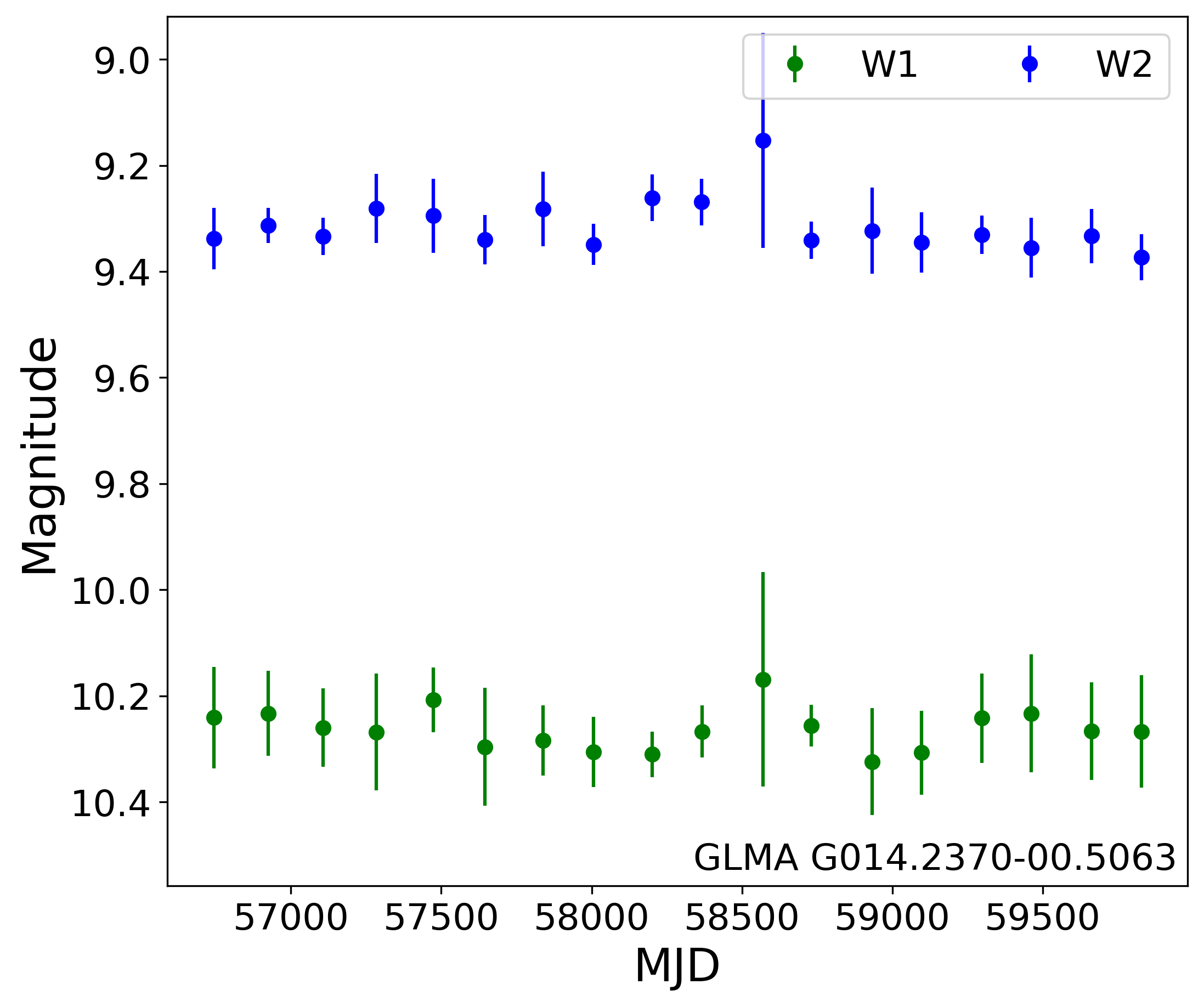}
    \end{minipage}    
    \caption{Representative light curves of several sources, observed in both the WISE W1 (green) and W2 (blue) bands.
    The source identifier is provided in the bottom right corner of each panel.}
    \label{fig:lc_mir_examples}
\end{figure*}

{These examples provide a guide for us to understand differences in assessing variability in W1 and W2.}
Differences in variability classification across the two mid-IR wavelengths may be due to either {sensitivity} issues, for instance extinction making one band much fainter than the other, or physical processes in which the variations have wavelength dependence, including emission line contamination for W2 \citep{Yoon2022}. 
To investigate these possibilities, we employ the Spearman rank correlation test to assess the correlation between individual source light curves at W1 and W2.
This non-parametric test was chosen due to its effectiveness in identifying monotonic relationships between two variables, making it suitable for analyzing light curve trends.
Our study involves a Spearman correlation analysis using both the initial observational data and outcomes from simulated datasets.

Initially, we apply the Spearman correlation test to obtain correlation coefficients and corresponding p-values to assess the presence of and significance of correlations between the W1 and W2 light curves.
Recognizing the potential influence of measurement uncertainties, we then extend our analysis to simulated datasets to further assess the reliability of these correlations.
To account for potential measurement uncertainties, we generate 100 simulated light curves for each source for both W1 and W2 bands.
Each simulated light curve is created by adding normally distributed random noise to the corresponding data points from the original light curve.
The standard deviation of this noise is set equal to the observational uncertainty associated with each original data point.
This method, implemented using \texttt{np.random.normal}, is designed to account for the potential impact of measurement errors on the correlation analysis. 
The Spearman correlation test is reapplied to each of these 100 simulated datasets for every source. We then compare the mean values of the correlation coefficients and p-values obtained from the simulations with those from the original dataset. This comparison offers insights into the stability and reliability of our results under the influence of observational uncertainties.

By analyzing both the observed and simulated data, we determine the correlation between the WISE W1 and W2 band light curves. The results are listed in Table~\ref{tab:variables_mir_sources}. 
For the 21 sources displaying robust measurable variations in both W1 and W2 bands, 19 had strong positive correlations and two had 
moderate-to-strong positive correlations.
Thus, despite the fact that sometimes these sources were classified differently by band, the light curves appear to be predominantly responding in the same manner, reflecting the same underlying physical conditions.


The 26 sources that showed variability in only one band present a more varied distribution of Spearman correlation agreement.
Ten sources were categorized as having a strong positive correlation, five as having a moderate-to-strong correlation, seven as having a weak-to-moderate correlation, and four as having no or very weak correlation.
For the fifteen sources, 58\%, exhibiting variability in only one WISE band (W1 or W2) and displaying a strong or moderate-to-strong positive correlation, it is expected that the same astrophysical phenomena are influencing both bands.
The variation in the other band might be too subtle or weak for our detection methods to identify.
This is also possible for the  7 sources, or 27\%, showing variability in only one band and exhibiting a weak-to-moderate Spearman correlation, though additional caution should be exercised for this small subset.
While the same physical phenomenon may affect both bands, the weak correlation could also indicate that the connection is not as direct or that other factors influence the observed variability.
These cases might require more in-depth analysis or additional data to clarify the nature of the variability.

Four mid-IR variable sources, all determined variable only at W2, have no or very weak correlation according to the Spearman correlation test results, suggesting that the observed flux variation may be specific to only one of the WISE bands.  
It is important to note, however, that noise levels in the data play a role in confusing the variability at W1 and should be taken into account when interpreting these results. Thus, each of these four variables was considered carefully by eye.
Among these four exceptional sources, only SPICY~81451 (Figure~\ref{fig:lc_mir_examples}) reveals a clear difference in variability between the W1 and W2 bands.
This finding warrants further investigation to understand the underlying mechanisms responsible for this distinct behavior
as discussed in further detail in Section~\ref{sec:disc_IR_interesting}.

\subsubsection{The confused case of SPICY 81451 in the mid-IR} \label{sec:disc_IR_interesting}

SPICY~81451 presents a complex and intriguing case. 
This source shows dramatic flux fluctuations in the WISE W2 band yet displays no discernible variation in the WISE W1 band (see Figure~\ref{fig:lc_mir_examples}).
A high angular resolution UKIDSS K image taken in 2012 \citep[0.7\arcsec;][]{lucas2008} reveals that there are three sources in proximity to the location of the WISE detection.
One of the three UKIDSS sources is classified as a galaxy (UGPS J181932.72-164510.2), a second source is considered stellar (UGPS J181933.04-164510.8), while the faintest source is considered a YSO (UGPS J181932.80-164511.9).
When examining the WISE peak positions, an 
offset of 2.7\arcsec\ is observed between the WISE W1 and W2.
Furthermore, comparing against the positions of the UKIDSS sources reveals that {the W1 and W2} WISE peaks coincide with a different UKIDSS source. 
Consequently, the W2 peak likely originates from the known YSO, while the W1 emission is likely associated with the nearby bright star.
This suggests that the discrepancy in W1 and W2 flux variability is likely due to the emission in each WISE band being dominated by different sources.
Notably, the YSO counterpart disappears in a more recent CFHT Ks image from 2020.
This 
is consistent with the decline observed in WISE W2 and supports the suggestion that {the two} WISE bands are dominated by emission for a  different source.


\subsection{Comparison with Variables in Low-mass Star-forming Regions}

Our survey of the \region\ region, which is forming stars of intermediate mass at a distance of $\sim 1.8$~kpc, complements the Gould Belt investigations by \citet{y.lee2021} and \citet{w.park2021}, which focus on low-mass star-forming regions within 500~pc.
Thus, a comparative analysis can offer valuable insights into the universality and diversity of star-forming processes.

In the sub-mm range, we identified only two robust sub-mm variable candidates out of 146 bright peaks (excluding the one robust variable extragalactic source). This corresponds to a lower detection rate compared to the 18 out of 83 protostars within the Gould Belt found by \citet{y.lee2021} over a similar timescale.
It is possible that this difference might be attributable to environmental factors, including an inherent weaker (or rarer) variability for the YSOs in the more massive \region. However, it is also worth recognizing the challenges of detecting sub-mm variability signatures at greater distances, where for a fixed JCMT beam the amount of envelope and nearby cloud contributing to the observed peak flux increases significantly and may therefore smother the localized variations. This latter possibility will be quantitatively explored  in a companion JCMT Transient Survey paper by Wang et al. (in preparation). 

Our mid-IR analysis of \region\ {used} 9 years of WISE monitoring to assess variability, compared to the 6.5 years of mid-IR Gould Belt monitoring employed by \citet{w.park2021}, who focused exclusively on the W2 band.
Considering only our W2 results for {Class 0/I and Class II}, we find consistency with \citet{w.park2021} in the relative numbers of variables by evolutionary class and variability type, with a higher fraction of {Class 0/I} than {Class II} being variable and longer secular timescales for the {Class 0/I} than {Class II}. However, the overall fraction of variability at both evolutionary stages is notably lower. This is  primarily due to a dearth of irregular variability in \region, which may be harder to discern for this sample since the typical mid-IR source is fainter than in the nearby Gould Belt. Setting aside the irregulars, the consistency in variability types between the Gould Belt and \region\ suggests a potential universality in the underlying mechanisms driving YSOs variability.

Thus, we have somewhat conflicting variability results at sub-mm and mid-IR between \region\ and the Gould Belt. However, accounting for expected complications monitoring at larger distances with fixed observing conditions, it is reasonable to assume that the underlying physical conditions responsible for stellar assembly remain similar across these star-forming environments. JCMT Transient Survey investigations of three additional massive star-forming regions will further probe this question.

\section{Summary} \label{sec:summ}

This study represents the first comprehensive attempt to characterize YSO variability in \region, a region poised for the formation of intermediate-mass stars. Variability across many years, at both sub-mm and mid-IR wavelengths has been investigated.

The sub-mm variability analysis uses 3.5 years of monthly monitoring \region\ with the JCMT and  identifies two robust YSO secular variables and one extragalactic source. Compared against previous sub-mm variability analyses in nearby low-mass star-forming regions within the Gould Belt, we find an order of magnitude decrease in detectable variability within \region. While this is possibly due to intrinsic differences in star formation activity with region complexity, it is most likely that observing inherent sub-mm variability is more difficult for sources at larger distances and in more complex environments.

The mid-IR variability analysis, using 9 years of bi-yearly WISE monitoring, reveals significant variability for 47 YSOs, with a notable fraction exhibiting variability in only one WISE band.
The comparison of variability types and the correlation analysis between the W1 and W2 bands suggests that for most YSOs the two light curves are intrinsically similar, with the majority of the differences in variability categorization being due to either complexity in the light curves, beyond the simple models being used to fit, and noise. Only for one source, SPICY~81451, are the two mid-IR light curves distinctly different, a consequence of 
two bright sources located within the WISE beam, with one source dominating at each wavelength.  

When compared with results from low-mass star-forming regions, our mid-IR results for \region\ suggest that some aspects of YSO variability are universal, including more variability at earlier stages and similar fractions of secular variability across regions. Potential regional differences are seen in the stochastic variability, with \region\ having fewer of these sources. However, this is potentially due to the larger distance to the region and the more crowded environment.  

Our study underscores the necessity of long-term, multi-wavelength monitoring campaigns across a variety of star formation environments to unravel the intricate processes governing the early evolution of YSOs.

\clearpage
\begin{acknowledgments}

This work was supported by the National Research Foundation of Korea through grants NRF-2020R1A6A3A01100208 \&  RS-2023-00242652 (G.P.). 
This work was partly supported by the Korea Astronomy and Space Science Institute grant funded by the Korea government(MSIT) (Project No. 2022-1-840-05).
D.J.\ is supported by NRC Canada and by an NSERC Discovery Grant.
J.-E. Lee is supported by the National Research Foundation of Korea (NRF) grant funded by the Korean government (MSIT) (grant number 2021R1A2C1011718).
GJH is supported by the National Key R\&D program
of China 2022YFA1603102 and by general grant 12173003 from the National Natural Science Foundation of China.

\end{acknowledgments}

\vspace{5mm}
\facilities{JCMT}

\software{Astropy \citep{astropy2013,astropy2018},
          TOPCAT \citep{Taylor2005}}

\appendix
\restartappendixnumbering

\section{Sub-mm Light Curves of Variable Candidates}

For completeness, Table~\ref{tab:all_submm} presents the locations, peak brightnesses, and variability measures used in this paper for all the sources found by the Fellwalker algorithm. 

\startlongtable
\begin{deluxetable}{rcrrr r}
\tabletypesize{\scriptsize}
\setlength{\tabcolsep}{2pt}
\tablecaption{Statistics of All Submillimeter Sources \label{tab:all_submm}}
\tablehead{
\colhead{\#} & \colhead{ID} & \colhead{$\mpflux$} & \colhead{SD} & \colhead{SD/SD$_{\rm fid}$} & \colhead{$|S/\Delta S|$}  \\
\colhead{} & \colhead{} & \colhead{($\mjybeam$)} & \colhead{($\mjybeam$)} & \colhead{} & \colhead{} 
}
\startdata
  1 & JCMTPP\_J181854.2$-$164754 &  14336.49 &    216.05 &      1.00 &      2.31 \\
  2 & JCMTPP\_J181812.2$-$164933 &   5409.12 &    125.06 &      1.52 &      5.92 \\
  3 & JCMTPP\_J181812.9$-$165727 &   3181.91 &     49.62 &      1.00 &      1.19 \\
  4 & JCMTPP\_J181811.2$-$165236 &   1047.20 &     19.46 &      0.92 &      1.17 \\
  5 & JCMTPP\_J181918.7$-$164357 &    915.80 &     25.16 &      1.28 &      0.33 \\
  6 & JCMTPP\_J181748.6$-$170030 &    794.43 &     17.59 &      0.96 &      1.25 \\
  7 & JCMTPP\_J181829.2$-$164454 &    757.06 &     15.68 &      0.87 &      0.54 \\
  8 & JCMTPP\_J181856.9$-$164612 &    738.71 &     16.18 &      0.91 &      1.13 \\
\enddata
\end{deluxetable}

\section{Variability Measurements from NEOWISE Light Curves for Final Sample YSOs}

We present the variability measures derived from NEOWISE light curves from the final sample of YSOs analyzed in our study, covering both the W1 and W2 bands.
Table~\ref{tab:variables_mir_sources} lists these measures from the 47 YSOs identified as exhibiting robust mid-IR variability across the six investigated types.
Table~\ref{tab:nonvariables_mir_sources} provides details for the remaining 132 YSOs in the final sample that did not exhibit detectable mid-IR variability.
For completenes, Figure~\ref{fig:lc_mir_remainder} shows mid-IR light curves for the 40 YSOs identified as exhibiting robust mid-IR variability, but not included in Figure~\ref{fig:lc_mir_examples}.

\begin{deluxetable*}{clccc ccccc ccccc c}
\tabletypesize{\scriptsize}
\setlength{\tabcolsep}{2pt}
\tablecaption{Properties of NEOWISE Light Curves for Potential Variable Mid-Infrared Sources \label{tab:variables_mir_sources}}
\tablehead{
\colhead{} & \colhead{} & \colhead{R.A.} & \colhead{Dec.} & \colhead{N} &
\colhead{$\mathrm{\Delta(W)}$} & \colhead{SD} &  \colhead{$\mathrm{\sigma}$} & 
\colhead{$S$\tablenotemark{\scriptsize b}} & 
\colhead{Period\tablenotemark{\scriptsize c}} &
\colhead{} & \colhead{} & \colhead{} & 
\colhead{} & \colhead{} & \colhead{} \\
\colhead{\#} & \colhead{ID\tablenotemark{a}} & \colhead{(deg)} & \colhead{(deg)} & \colhead{} & 
\colhead{(mag)} & \colhead{(mag)} & \colhead{(mag)} &
\colhead{(\%/yr)} & \colhead{(yr)} & 
\colhead{$\mathrm{FAP_{Lin}}$} & \colhead{$\mathrm{FAP_{LSP}}$} & 
\colhead{Type\tablenotemark{\scriptsize d}} & 
\colhead{Class\tablenotemark{\scriptsize e}} & \colhead{Outside\tablenotemark{\scriptsize f}} &
\colhead{Corr.\tablenotemark{\scriptsize g}} 
}
\startdata
  1 & SPICY 79195      & 274.3010 & $-$16.9586 & \makecell{ 18 \\ 18 } & \makecell{ 0.310 \\ 0.452 } & \makecell{ 0.084 \\ 0.106 } & \makecell{ 0.049 \\ 0.035 } & \makecell{ \nodata \\ \nodata } & \makecell{ \nodata \\ \nodata } & \makecell{ 7.4e$-$01 \\ 6.7e$-$01 } & \makecell{ 6.7e$-$02 \\ 6.8e$-$02 } & \makecell{ \nodata  \\ irregular } & 0/I& \nodata  & C \\ \hline
  2 & SPICY 79250      & 274.3144 & $-$16.8679 & \makecell{ 18 \\ 18 } & \makecell{ 1.338 \\ 1.057 } & \makecell{ 0.438 \\ 0.370 } & \makecell{ 0.109 \\ 0.125 } & \makecell{ \nodata \\ \nodata } & \makecell{ 1.496   \\ 1.496   } & \makecell{ 9.0e$-$01 \\ 9.5e$-$01 } & \makecell{ 6.5e$-$04 \\ 6.1e$-$07 } & \makecell{ periodic \\ periodic } &III & Y & A \\ \hline
  3 & SPICY 79296      & 274.3211 & $-$16.9943 & \makecell{ 18 \\ 18 } & \makecell{ 0.288 \\ 0.266 } & \makecell{ 0.079 \\ 0.071 } & \makecell{ 0.055 \\ 0.122 } & \makecell{ $-$2.798 \\ \nodata } & \makecell{ 10.021  \\ \nodata } & \makecell{ 8.0e$-$03 \\ 2.6e$-$01 } & \makecell{ 2.4e$-$03 \\ 2.7e$-$02 } & \makecell{ curved   \\ \nodata  } & 0/I& \nodata & C \\ \hline
  4 & SPICY 79425      & 274.3571 & $-$17.0367 & \makecell{ 18 \\ 18 } & \makecell{ 4.440 \\ 3.334 } & \makecell{ 1.183 \\ 0.921 } & \makecell{ 0.195 \\ 0.084 } & \makecell{ \nodata \\ +10.315 } & \makecell{ \nodata \\ 6.795   } & \makecell{ 5.4e$-$01 \\ 1.2e$-$01 } & \makecell{ 5.0e$-$02 \\ 1.5e$-$04 } & \makecell{ irregular \\ curved   } & 0/I& \nodata & A \\ \hline
  5 & SPICY 79472      & 274.3713 & $-$17.0259 & \makecell{ 18 \\ 18 } & \makecell{ 0.322 \\ 0.406 } & \makecell{ 0.087 \\ 0.099 } & \makecell{ 0.088 \\ 0.070 } & \makecell{ \nodata \\ $-$2.617 } & \makecell{ \nodata \\ 13.142  } & \makecell{ 8.6e$-$03 \\ 1.2e$-$02 } & \makecell{ 3.1e$-$02 \\ 7.5e$-$03 } & \makecell{ \nodata  \\ curved   } & II & \nodata & A \\
\enddata
\tablecomments{Continued on next page}
\end{deluxetable*}

\addtocounter{table}{-1}
\begin{deluxetable*}{ccccc ccccc ccccc c}[!]
\tabletypesize{\scriptsize}
\setlength{\tabcolsep}{2pt}
\tablecaption{Continued from previous page}
\tablehead{
\colhead{} & \colhead{} & \colhead{R.A.} & \colhead{Dec.} & \colhead{N} &
\colhead{$\mathrm{\Delta(W)}$} & \colhead{SD} &  \colhead{$\mathrm{\sigma}$} & 
\colhead{$S$\tablenotemark{\scriptsize b}} & 
\colhead{Period\tablenotemark{\scriptsize c}} &
\colhead{} & \colhead{} & \colhead{} & 
\colhead{} & \colhead{} & \colhead{} \\
\colhead{\#} & \colhead{ID\tablenotemark{\scriptsize a}} & \colhead{(deg)} & \colhead{(deg)} & \colhead{} & 
\colhead{(mag)} & \colhead{(mag)} & \colhead{(mag)} &
\colhead{(\%/yr)} & \colhead{(yr)} & 
\colhead{$\mathrm{FAP_{Lin}}$} & \colhead{$\mathrm{FAP_{LSP}}$} & 
\colhead{Type\tablenotemark{\scriptsize d}} & 
\colhead{Class\tablenotemark{\scriptsize e}} & \colhead{Outside\tablenotemark{\scriptsize f}} &
\colhead{Corr.\tablenotemark{\scriptsize g}} 
}
\startdata
 36 & SPICY 81140      & 274.7921 & $-$16.6523 & \makecell{ 18 \\ 18 } & \makecell{ 0.751 \\ 0.695 } & \makecell{ 0.251 \\ 0.242 } & \makecell{ 0.077 \\ 0.051 } & \makecell{+6.771 \\ +10.444 } & \makecell{ 10.021  \\ 10.021  } & \makecell{ 4.9e$-$03 \\ 1.8e$-$03 } & \makecell{ 3.0e$-$04 \\ 5.8e$-$04 } & \makecell{ curved   \\ curved   } & II & \nodata & A \\ \hline
 37 & SPICY 81164      & 274.7980 & $-$16.8000 & \makecell{ 18 \\ 18 } & \makecell{ 0.246 \\ 0.230 } & \makecell{ 0.072 \\ 0.069 } & \makecell{ 0.102 \\ 0.059 } & \makecell{ \nodata \\ $-$0.138 } & \makecell{ \nodata \\ 8.099   } & \makecell{ 1.1e$-$01 \\ 9.9e$-$01 } & \makecell{ 4.4e$-$04 \\ 4.9e$-$04 } & \makecell{ \nodata  \\ curved   } & II & \nodata & C\tablenotemark{\scriptsize h} \\ \hline
 38 & SPICY 81226      & 274.8124 & $-$16.8308 & \makecell{ 18 \\ 18 } & \makecell{ 0.395 \\ 0.417 } & \makecell{ 0.108 \\ 0.102 } & \makecell{ 0.237 \\ 0.050 } & \makecell{ \nodata \\ +5.404 } & \makecell{ \nodata \\ \nodata } & \makecell{ 1.6e$-$01 \\ 4.7e$-$06 } & \makecell{ 5.5e$-$01 \\ 1.3e$-$03 } & \makecell{ \nodata  \\ linear   } & 0/I& \nodata & D \\ \hline
 39 & SPICY 81374      & 274.8585 & $-$17.0125 & \makecell{ 18 \\ 18 } & \makecell{ 0.566 \\ 0.515 } & \makecell{ 0.151 \\ 0.126 } & \makecell{ 0.092 \\ 0.047 } & \makecell{ \nodata \\ $-$1.994 } & \makecell{ \nodata \\ 6.794   } & \makecell{ 6.7e$-$01 \\ 6.1e$-$01 } & \makecell{ 3.9e$-$02 \\ 6.5e$-$03 } & \makecell{ \nodata  \\ curved   } & 0/I& Y & A \\ \hline
 40 & SPICY 81450      & 274.8866 & $-$16.7596 & \makecell{ 18 \\ 18 } & \makecell{ 0.880 \\ 0.799 } & \makecell{ 0.246 \\ 0.220 } & \makecell{ 0.058 \\ 0.049 } & \makecell{ \nodata \\ \nodata } & \makecell{ \nodata \\ \nodata } & \makecell{ 5.3e$-$03 \\ 4.7e$-$02 } & \makecell{ 3.8e$-$02 \\ 2.9e$-$02 } & \makecell{ irregular \\ irregular } & II & \nodata & A \\
\enddata
\tablecomments{In the table, where two values are stacked vertically, the top value corresponds to W1 while the bottom value denotes W2.}
\tablenotetext{a}{All source identifiers (IDs) are based on designations from two key references: \citet{kuhn2021} and \citet{povich2010}. \citet{kuhn2021} utilizes the SPICY (Spitzer/IRAC Candidate YSO) numbering system, while \citet{povich2010} refer to sources by their GLIMPSE Archive (GLMA) coordinates. SPICY IDs are numerical, whereas GLMA IDs are presented in a Galactic coordinate format.}
\tablenotetext{b}{Best-fit linear slope.} 
\tablenotetext{c}{LSP period for periodic or curved sources.}
\tablenotetext{d}{Variable type.}
\tablenotetext{e}{YSO classification.}
\tablenotetext{f}{All sources are presumed to be associated with \region, unless marked as 'outside' (Y) in accordance with the determinations by \citet{kuhn2021}.}
\tablenotetext{g}{The Spearman correlation between WISE W1 and W2 light curves.
categories are as follows: A = strong positive correlation, B = moderate$-$to-strong positive correlation, C = weak-to-moderate positive correlation, D = No or very weak correlation.}
\tablenotetext{h}{Those who show notable differences between the original and simulated data.}
\end{deluxetable*}

\begin{deluxetable*}{ccccc ccccc cc}
\tabletypesize{\scriptsize}
\setlength{\tabcolsep}{2pt}
\tablecaption{Statistics of NEOWISE Light Curves for Non-Variable Mid-Infrared Sources \label{tab:nonvariables_mir_sources}}
\tablehead{
\colhead{} & \colhead{} & \colhead{R.A.} & \colhead{Dec.} & \colhead{N} &
\colhead{$\mathrm{\Delta(W)}$} & \colhead{SD} &  \colhead{$\mathrm{\sigma}$} & 
\colhead{} & \colhead{} & 
\colhead{} & \colhead{}  \\
\colhead{\#} & \colhead{ID\tablenotemark{a}} & \colhead{(deg)} & \colhead{(deg)} & \colhead{} & 
\colhead{(mag)} & \colhead{(mag)} & \colhead{(mag)} & 
\colhead{$\mathrm{FAP_{Lin}}$} & \colhead{$\mathrm{FAP_{LSP}}$} & 
\colhead{Class\tablenotemark{\scriptsize b}} & \colhead{Outside\tablenotemark{\scriptsize c}} 
}
\startdata
  1 & SPICY 79255      & 274.3150 & $-$16.9375 & \makecell{ 17 \\ 17 } & \makecell{ 0.905 \\ 0.313 } & \makecell{ 0.216 \\ 0.083 } & \makecell{ 0.205 \\ 0.196 } & \makecell{ 7.5e$-$01 \\ 2.5e$-$01 } & \makecell{ 5.2e$-$01 \\ 5.3e$-$01 } &II& \nodata \\ \hline
  2 & SPICY 79278      & 274.3184 & $-$16.9781 & \makecell{ 18 \\ 18 } & \makecell{ 0.282 \\ 0.317 } & \makecell{ 0.066 \\ 0.067 } & \makecell{ 0.045 \\ 0.047 } & \makecell{ 9.6e$-$01 \\ 7.4e$-$01 } & \makecell{ 5.1e$-$01 \\ 3.7e$-$01 } &II& \nodata \\ \hline
  3 & SPICY 79352      & 274.3393 & $-$16.8787 & \makecell{ 18 \\ 18 } & \makecell{ 0.283 \\ 0.443 } & \makecell{ 0.071 \\ 0.107 } & \makecell{ 0.105 \\ 0.064 } & \makecell{ 4.6e$-$01 \\ 7.1e$-$02 } & \makecell{ 7.9e$-$02 \\ 4.7e$-$02 } &0/I& \nodata \\ \hline
  4 & SPICY 79376      & 274.3441 & $-$16.9843 & \makecell{ 18 \\ 18 } & \makecell{ 0.368 \\ 0.126 } & \makecell{ 0.106 \\ 0.030 } & \makecell{ 0.203 \\ 0.100 } & \makecell{ 5.4e$-$01 \\ 8.7e$-$01 } & \makecell{ 3.2e$-$01 \\ 8.4e$-$01 } &0/I& \nodata \\ \hline
  5 & SPICY 79423      & 274.3565 & $-$17.0686 & \makecell{ 16 \\ 16 } & \makecell{ 0.743 \\ 0.286 } & \makecell{ 0.176 \\ 0.076 } & \makecell{ 0.195 \\ 0.148 } & \makecell{ 9.5e$-$01 \\ 9.1e$-$01 } & \makecell{ 5.0e$-$01 \\ 2.7e$-$01 } &II& \nodata \\ \hline
  6 & SPICY 79441      & 274.3617 & $-$17.0703 & \makecell{ 18 \\ 18 } & \makecell{ 0.282 \\ 0.127 } & \makecell{ 0.071 \\ 0.032 } & \makecell{ 0.158 \\ 0.073 } & \makecell{ 9.5e$-$01 \\ 4.8e$-$01 } & \makecell{ 8.2e$-$02 \\ 4.0e$-$01 } &0/I& \nodata \\ \hline
  7 & SPICY 79452      & 274.3648 & $-$17.0273 & \makecell{ 18 \\ 18 } & \makecell{ 0.594 \\ 0.378 } & \makecell{ 0.147 \\ 0.103 } & \makecell{ 0.236 \\ 0.107 } & \makecell{ 6.1e$-$02 \\ 4.2e$-$01 } & \makecell{ 4.5e$-$01 \\ 4.1e$-$01 } &0/I& \nodata \\
\enddata
\tablecomments{Continued on next page}
\end{deluxetable*}

\addtocounter{table}{-1}
\begin{deluxetable*}{ccccc ccccc cc}
\tabletypesize{\scriptsize}
\setlength{\tabcolsep}{2pt}
\tablecaption{Continued from previous page}
\tablehead{
\colhead{} & \colhead{} & \colhead{R.A.} & \colhead{Dec.} & \colhead{N} &
\colhead{$\mathrm{\Delta(W)}$} & \colhead{SD} &  \colhead{$\mathrm{\sigma}$} & 
\colhead{} & \colhead{} & 
\colhead{} & \colhead{}  \\
\colhead{\#} & \colhead{ID\tablenotemark{a}} & \colhead{(deg)} & \colhead{(deg)} & \colhead{} & 
\colhead{(mag)} & \colhead{(mag)} & \colhead{(mag)} & 
\colhead{$\mathrm{FAP_{Lin}}$} & \colhead{$\mathrm{FAP_{LSP}}$} & 
\colhead{Class\tablenotemark{\scriptsize b}} & \colhead{Outside\tablenotemark{\scriptsize c}} 
}
\startdata
 36 & SPICY 80170      & 274.5523 & $-$16.8179 & \makecell{ 18 \\ 18 } & \makecell{ 0.374 \\ 0.395 } & \makecell{ 0.079 \\ 0.082 } & \makecell{ 0.060 \\ 0.072 } & \makecell{ 8.5e$-$01 \\ 8.4e$-$01 } & \makecell{ 1.7e$-$02 \\ 1.2e$-$02 } &II& \nodata \\ \hline
 37 & SPICY 80173      & 274.5530 & $-$16.8240 & \makecell{ 17 \\ 17 } & \makecell{ 0.619 \\ 0.445 } & \makecell{ 0.167 \\ 0.114 } & \makecell{ 0.171 \\ 0.099 } & \makecell{ 3.1e$-$01 \\ 4.0e$-$01 } & \makecell{ 1.9e$-$02 \\ 3.2e$-$01 } &II& \nodata \\ \hline
 38 & SPICY 80175      & 274.5531 & $-$16.8173 & \makecell{ 18 \\ 18 } & \makecell{ 0.333 \\ 0.353 } & \makecell{ 0.077 \\ 0.078 } & \makecell{ 0.065 \\ 0.069 } & \makecell{ 1.0e+00 \\ 1.0e+00 } & \makecell{ 3.1e$-$01 \\ 2.0e$-$01 } & uncertain & \nodata \\ \hline
 39 & SPICY 80203      & 274.5607 & $-$16.7960 & \makecell{ 18 \\ 18 } & \makecell{ 0.306 \\ 0.207 } & \makecell{ 0.078 \\ 0.048 } & \makecell{ 0.128 \\ 0.074 } & \makecell{ 2.7e$-$01 \\ 4.1e$-$01 } & \makecell{ 7.5e$-$01 \\ 3.8e$-$01 } &II& \nodata \\ \hline
 40 & SPICY 80214      & 274.5635 & $-$16.8258 & \makecell{ 18 \\ 18 } & \makecell{ 0.218 \\ 0.239 } & \makecell{ 0.055 \\ 0.074 } & \makecell{ 0.131 \\ 0.085 } & \makecell{ 5.7e$-$02 \\ 9.7e$-$01 } & \makecell{ 2.4e$-$02 \\ 2.0e$-$02 } &0/I& \nodata \\ \hline
 41 & SPICY 80255      & 274.5753 & $-$16.7754 & \makecell{ 18 \\ 18 } & \makecell{ 0.233 \\ 0.080 } & \makecell{ 0.054 \\ 0.026 } & \makecell{ 0.113 \\ 0.054 } & \makecell{ 7.6e$-$01 \\ 1.3e$-$02 } & \makecell{ 3.4e$-$02 \\ 8.3e$-$03 } &II& \nodata \\ \hline
 42 & SPICY 80277      & 274.5795 & $-$16.9806 & \makecell{ 18 \\ 18 } & \makecell{ 0.576 \\ 0.667 } & \makecell{ 0.155 \\ 0.172 } & \makecell{ 0.148 \\ 0.065 } & \makecell{ 1.3e$-$01 \\ 4.5e$-$03 } & \makecell{ 4.9e$-$01 \\ 3.4e$-$02 } &II& \nodata \\
\enddata
\tablecomments{Continued on next page}
\end{deluxetable*}

\addtocounter{table}{-1}
\begin{deluxetable*}{ccccc ccccc cc}
\tabletypesize{\scriptsize}
\setlength{\tabcolsep}{2pt}
\tablecaption{Continued from previous page}
\tablehead{
\colhead{} & \colhead{} & \colhead{R.A.} & \colhead{Dec.} & \colhead{N} &
\colhead{$\mathrm{\Delta(W)}$} & \colhead{SD} &  \colhead{$\mathrm{\sigma}$} & 
\colhead{} & \colhead{} & 
\colhead{} & \colhead{}  \\
\colhead{\#} & \colhead{ID\tablenotemark{a}} & \colhead{(deg)} & \colhead{(deg)} & \colhead{} & 
\colhead{(mag)} & \colhead{(mag)} & \colhead{(mag)} & 
\colhead{$\mathrm{FAP_{Lin}}$} & \colhead{$\mathrm{FAP_{LSP}}$} & 
\colhead{Class\tablenotemark{\scriptsize b}} & \colhead{Outside\tablenotemark{\scriptsize c}} 
}
\startdata
 71 & SPICY 80749      & 274.6888 & $-$16.5462 & \makecell{ 18 \\ 18 } & \makecell{ 0.308 \\ 0.422 } & \makecell{ 0.079 \\ 0.108 } & \makecell{ 0.069 \\ 0.059 } & \makecell{ 6.2e$-$01 \\ 1.0e+00 } & \makecell{ 3.8e$-$02 \\ 3.8e$-$02 } &0/I& Y \\ \hline
 72 & SPICY 80814      & 274.7041 & $-$16.8393 & \makecell{ 14 \\ 14 } & \makecell{ 0.765 \\ 0.937 } & \makecell{ 0.258 \\ 0.226 } & \makecell{ 0.168 \\ 0.174 } & \makecell{ 5.9e$-$01 \\ 6.8e$-$01 } & \makecell{ 2.5e$-$02 \\ 3.7e$-$02 } &II& \nodata \\ \hline
 73 & SPICY 80835      & 274.7117 & $-$16.8712 & \makecell{ 16 \\ 16 } & \makecell{ 0.717 \\ 0.305 } & \makecell{ 0.146 \\ 0.090 } & \makecell{ 0.166 \\ 0.111 } & \makecell{ 4.1e$-$01 \\ 5.3e$-$01 } & \makecell{ 7.6e$-$01 \\ 4.5e$-$01 } &II& \nodata \\ \hline
 74 & SPICY 80873      & 274.7209 & $-$16.6357 & \makecell{ 18 \\ 18 } & \makecell{ 0.260 \\ 0.303 } & \makecell{ 0.063 \\ 0.085 } & \makecell{ 0.143 \\ 0.119 } & \makecell{ 6.3e$-$01 \\ 3.2e$-$01 } & \makecell{ 1.9e$-$02 \\ 1.2e$-$02 } &II& \nodata \\ \hline
 75 & SPICY 80883      & 274.7232 & $-$16.6883 & \makecell{ 18 \\ 18 } & \makecell{ 0.286 \\ 0.189 } & \makecell{ 0.071 \\ 0.047 } & \makecell{ 0.138 \\ 0.076 } & \makecell{ 9.2e$-$01 \\ 4.4e$-$01 } & \makecell{ 1.4e$-$01 \\ 2.0e$-$01 } &II& \nodata \\ \hline
 76 & SPICY 80894      & 274.7273 & $-$16.7800 & \makecell{ 18 \\ 18 } & \makecell{ 0.274 \\ 0.133 } & \makecell{ 0.059 \\ 0.033 } & \makecell{ 0.058 \\ 0.048 } & \makecell{ 9.2e$-$01 \\ 8.0e$-$01 } & \makecell{ 4.1e$-$01 \\ 9.3e$-$01 } &II& \nodata \\ \hline
 77 & SPICY 80919      & 274.7329 & $-$16.8060 & \makecell{ 13 \\ 13 } & \makecell{ 0.690 \\ 0.478 } & \makecell{ 0.196 \\ 0.128 } & \makecell{ 0.246 \\ 0.148 } & \makecell{ 5.9e$-$01 \\ 5.4e$-$01 } & \makecell{ 1.2e$-$01 \\ 4.8e$-$01 } &II& \nodata \\
\enddata
\tablecomments{Continued on next page}
\end{deluxetable*}

\addtocounter{table}{-1}
\begin{deluxetable*}{ccccc ccccc cc}
\tabletypesize{\scriptsize}
\setlength{\tabcolsep}{2pt}
\tablecaption{Continued from previous page}
\tablehead{
\colhead{} & \colhead{} & \colhead{R.A.} & \colhead{Dec.} & \colhead{N} &
\colhead{$\mathrm{\Delta(W)}$} & \colhead{SD} &  \colhead{$\mathrm{\sigma}$} & 
\colhead{} & \colhead{} & 
\colhead{} & \colhead{}  \\
\colhead{\#} & \colhead{ID\tablenotemark{a}} & \colhead{(deg)} & \colhead{(deg)} & \colhead{} & 
\colhead{(mag)} & \colhead{(mag)} & \colhead{(mag)} & 
\colhead{$\mathrm{FAP_{Lin}}$} & \colhead{$\mathrm{FAP_{LSP}}$} & 
\colhead{Class\tablenotemark{\scriptsize b}} & \colhead{Outside\tablenotemark{\scriptsize c}} 
}
\startdata
106 & SPICY 81254      & 274.8186 & $-$16.7659 & \makecell{ 18 \\ 18 } & \makecell{ 0.325 \\ 0.389 } & \makecell{ 0.090 \\ 0.091 } & \makecell{ 0.129 \\ 0.086 } & \makecell{ 7.9e$-$01 \\ 5.8e$-$01 } & \makecell{ 7.5e$-$02 \\ 2.6e$-$01 } & III & \nodata \\ \hline
107 & SPICY 81255      & 274.8191 & $-$16.7443 & \makecell{ 18 \\ 18 } & \makecell{ 0.257 \\ 0.301 } & \makecell{ 0.065 \\ 0.074 } & \makecell{ 0.070 \\ 0.045 } & \makecell{ 7.5e$-$01 \\ 8.7e$-$01 } & \makecell{ 4.2e$-$02 \\ 1.6e$-$01 } &II& \nodata \\ \hline
108 & SPICY 81261      & 274.8227 & $-$16.6042 & \makecell{ 18 \\ 18 } & \makecell{ 0.250 \\ 0.226 } & \makecell{ 0.061 \\ 0.059 } & \makecell{ 0.131 \\ 0.112 } & \makecell{ 3.6e$-$01 \\ 6.4e$-$01 } & \makecell{ 7.5e$-$01 \\ 9.5e$-$02 } &II& \nodata \\ \hline
109 & SPICY 81298      & 274.8349 & $-$16.8103 & \makecell{ 18 \\ 18 } & \makecell{ 0.371 \\ 0.181 } & \makecell{ 0.111 \\ 0.051 } & \makecell{ 0.148 \\ 0.070 } & \makecell{ 7.9e$-$02 \\ 5.2e$-$02 } & \makecell{ 2.5e$-$01 \\ 7.4e$-$01 } &II& \nodata \\ \hline
110 & SPICY 81306      & 274.8366 & $-$16.7339 & \makecell{ 18 \\ 18 } & \makecell{ 0.523 \\ 0.641 } & \makecell{ 0.170 \\ 0.191 } & \makecell{ 0.181 \\ 0.135 } & \makecell{ 9.8e$-$01 \\ 9.1e$-$01 } & \makecell{ 2.4e$-$02 \\ 4.1e$-$01 } &II& \nodata \\ \hline
111 & SPICY 81331      & 274.8429 & $-$16.7194 & \makecell{ 16 \\ 16 } & \makecell{ 0.418 \\ 0.401 } & \makecell{ 0.116 \\ 0.109 } & \makecell{ 0.177 \\ 0.157 } & \makecell{ 6.1e$-$01 \\ 1.2e$-$01 } & \makecell{ 3.2e$-$03 \\ 1.2e$-$02 } &II& \nodata \\ \hline
112 & SPICY 81332      & 274.8433 & $-$16.7227 & \makecell{ 18 \\ 18 } & \makecell{ 0.267 \\ 0.202 } & \makecell{ 0.080 \\ 0.068 } & \makecell{ 0.134 \\ 0.106 } & \makecell{ 9.7e$-$01 \\ 2.0e$-$01 } & \makecell{ 8.4e$-$02 \\ 1.9e$-$02 } &0/I& \nodata \\
\enddata
\tablecomments{In the table, where two values are stacked vertically, the top value corresponds to W1 while the bottom value denotes W2.}
\tablenotetext{a}{All source identifiers (IDs) are based on designations from two key references: \citet{kuhn2021} and \citet{povich2010}. \citet{kuhn2021} utilizes the SPICY (Spitzer/IRAC Candidate YSO) numbering system, while \citet{povich2010} refer to sources by their GLIMPSE Archive (GLMA) coordinates. SPICY IDs are numerical, whereas GLMA IDs are presented in a Galactic coordinate format.}
\tablenotetext{b}{YSO classification.}
\tablenotetext{c}{All sources are presumed to be associated with \region, unless marked as 'outside' (Y) in accordance with the determinations by \citet{kuhn2021}.}
\end{deluxetable*}

\begin{figure*}
    \centering
    \includegraphics[width=.97\textwidth]{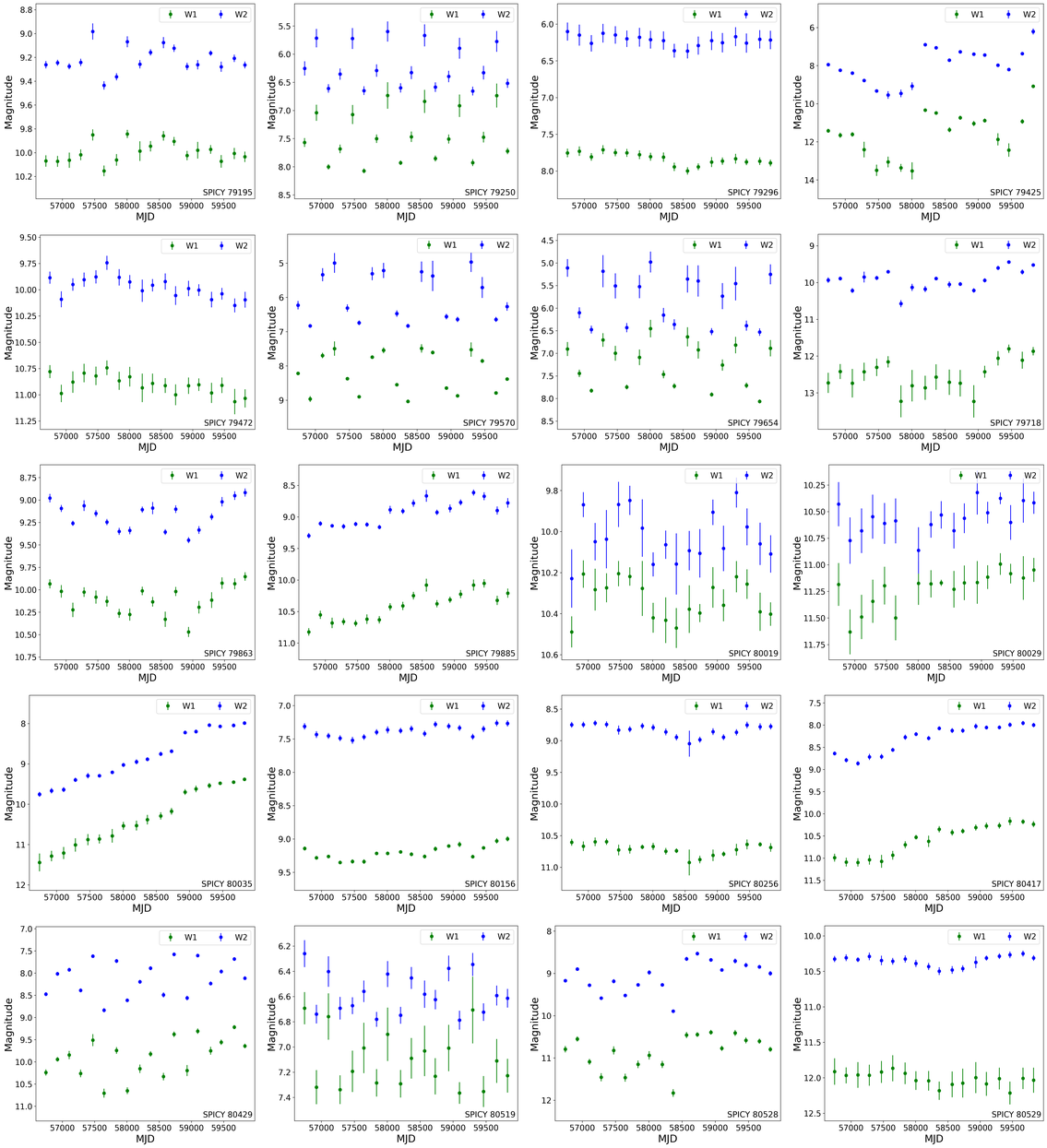}
    \caption{Mid-IR light curves for sources in the final WISE sample, included in either the W1 or W2 analysis, are presented.
    The total number of sources is 179.
    Sources shown in Figure~\ref{fig:lc_mir_examples} are not included here.
    The first 40 sources exhibit flux variability in either the W1 or W2 band.
    The subsequent panels display sources that were not observed to vary in this study.
    The source identifier is provided in the bottom right corner of each panel.}
    \label{fig:lc_mir_remainder}
\end{figure*}

\begin{figure*}
    \centering
    \addtocounter{figure}{-1}
    \includegraphics[width=.97\textwidth]{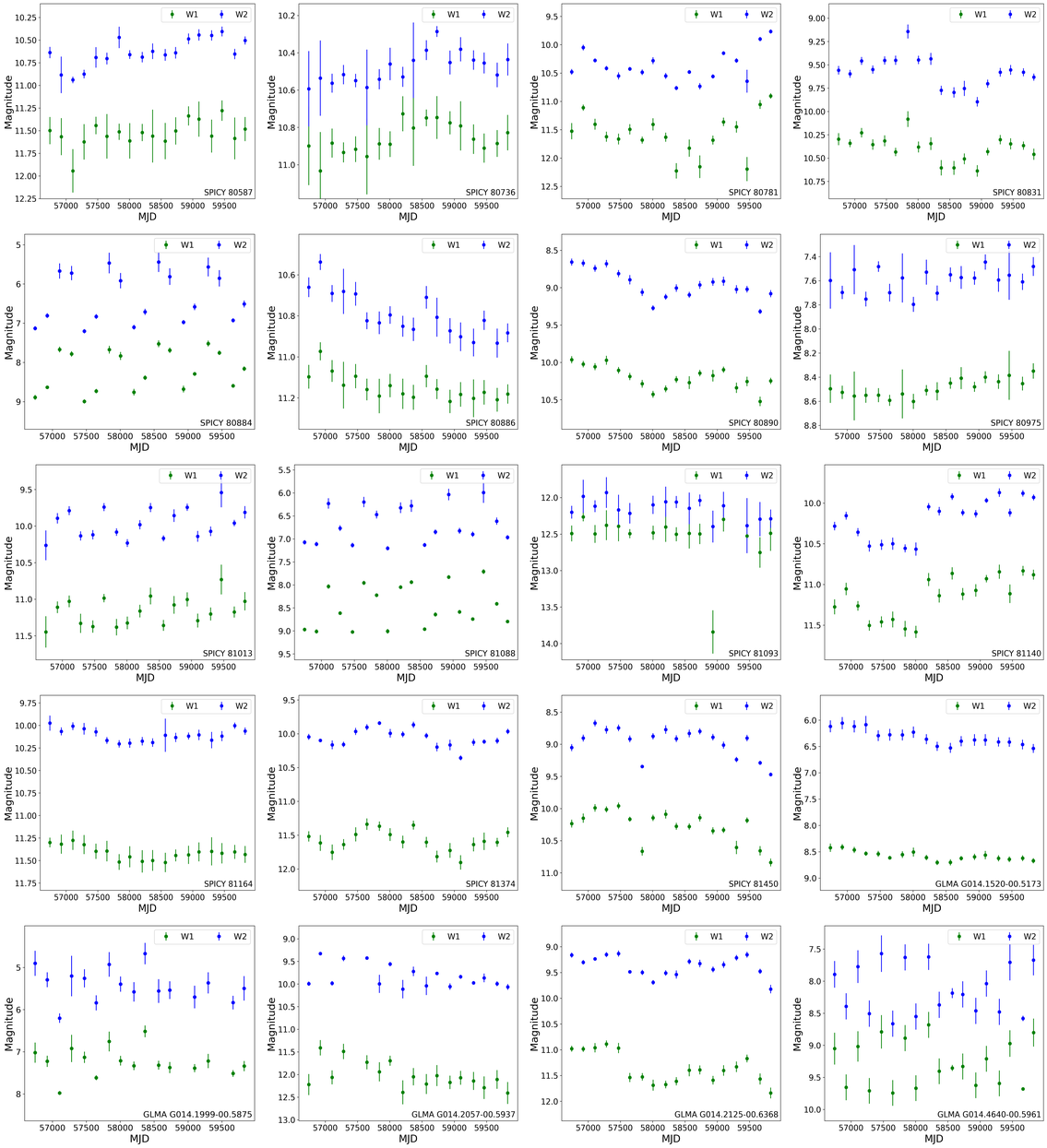}
    \caption{Continued from the previous page.}
\end{figure*}

\bibliography{main}{}
\bibliographystyle{aasjournal}

\end{document}